\documentclass[aip,jcp,reprint,amsmath,amssymb,nobalancelastpage,floatfix]{revtex4-2}
\usepackage{graphicx}
\usepackage{xcolor}
\usepackage{dblfloatfix}
\usepackage{wrapfig}
\usepackage{placeins}
\usepackage{tcolorbox}
\usepackage[suffix=]{epstopdf}
\usepackage[utf8]{inputenc}
\renewcommand{\vec}[1]
{\boldsymbol{\mathrm{#1}}}
\usepackage[absolute,overlay]{textpos}
\graphicspath{{./images/}}
\usepackage[caption=false]{subfig}

\usepackage[hidelinks]{hyperref}
\usepackage{xr-hyper}
\renewcommand{\selectlanguage}[1]{}

\usepackage{titlesec}
\titlespacing*{\section}{0pt}{1.8ex plus 1ex minus .2ex}{1.0ex plus .2ex}
\titlespacing*{\subsection}{0pt}{1.8ex plus 1ex minus .2ex}{1.0ex plus .2ex}

\makeatletter
\newcommand\etc{etc\@ifnextchar.{}{.\@}}
\newcommand\etal{et al\@ifnextchar.{}{.\@}}
\newcommand\eg{e.\,g\,\@ifnextchar.{}{.\@}}
\makeatother

\DeclareUnicodeCharacter{0308}{\"{}}
\DeclareUnicodeCharacter{2009}{\,}

\begin{document}
\title{Effect of Pre-Shear and Dispersity on Crystallization of a Model Polymer with Soft Pair Interactions using Molecular Dynamics Simulations}

\author{Tzortzis Koulaxizis}
\affiliation{Department of Chemical and Biomolecular Engineering, University of Illinois, Urbana--Champaign, Illinois 61801, USA}

\author{Antonia Statt}
\email{statt@illinois.edu}
\affiliation{Department of Materials Science and Engineering, The Grainger College of Engineering, University of Illinois Urbana-Champaign, Urbana, Illinois 61801, USA}

\begin{abstract}
\begin{wrapfigure}[7]{r}{0.30\linewidth} 
  \centering
  \vspace{-18pt}
  \hspace{-90pt} 
  \includegraphics[width=1.0\linewidth]{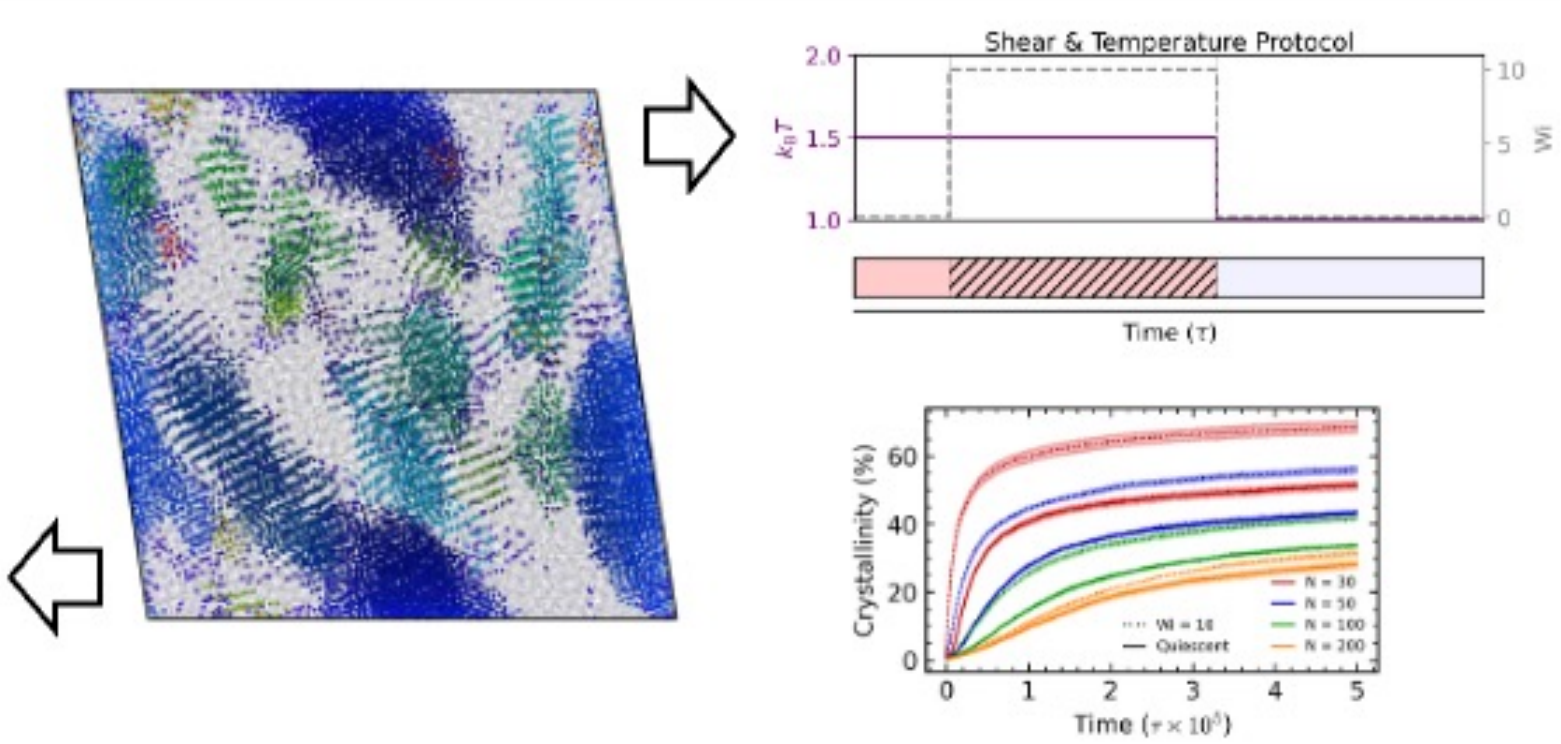} 
  \vspace{-5pt} 
\end{wrapfigure}
Polymer crystallization is a process of great interest in both fundamental theory and industrial settings, particularly in polymer processing and applications involving semi-crystalline materials. The effect of processing on the initial stages of crystallization is not fully understood. Our study investigates the influence of pre-shear on monodisperse melts and bidisperse blends of a generic, segmentally coarse-grained polymer model. Through molecular dynamics simulations, we explore how polydispersity affects crystallization, where we found that the addition of short chains to a melt of longer chains increased the final crystallinity by about 10\%, and increased the initial growth rate by roughly a factor of two. 
In contrast, however, pre-shearing the hot melt before quenching only showed a minor increase in both growth rates and final crystallinty, except in monodisperse melts of short chains. Crystal grain shapes were most influenced by pre-shearing monodisperse melts, where both asphericity  and prolateness decreased. 
Additionally, we determined topological connectivity of crystal grains through tie- and loop-chain analysis. Again, only monodisperse melts showed a significant increase of tie chain fractions with pre-shear, while all other systems showed only modest increases.
Our findings provide insight into the changes of crystallinity and cluster morphologies that emerge when pre-sheared, offering a deeper understanding of the initial crystallization processes in polymer melts when subjected to pre-shear.
\end{abstract}

\maketitle

\section{Introduction}
Semicrystalline polymers, in which the distribution and connectivity of amorphous and crystalline regions determine their mechanical and thermal performance, are important for commodity materials.\cite{wang2016flow} Here, polymer crystallization is typically directly impacted by processing \cite{graham_understanding_2019,chandran2019processing}, as shear and extensional flows are intrinsic to industrial operations. Processes such as injection molding, extrusion, fiber spinning, and film blowing require  polymer flow, simultaneously promoting polymer alignment and crystallization, thereby influencing final morphology and performance\cite{kim2005flow,continuous_crystallizers, pantani2005modeling,doufas2001simulation,larrondo1981electrostatic}. 

Several mechanistic frameworks have been proposed to describe polymer crystallization. In the classical picture, crystallization proceeds through nucleation and subsequent growth \cite{graham_modelling_2014,piorkowska2013handbook,cui_nonequilibrium_2015,xu_concepts_2021}, whereas alternative descriptions involve cooperative ordering \cite{semiflexible_kawak, strobl_colloquium_2009,dtree_kawak_oligomer} or multi-step pathways involving transient partially ordered intermediates \cite{hamad_lifetime_2015,luo_growth_2011,azzurri_insights_2008,azzurri_lifetime_2005,cavallo_flow_2010, cui_multiscale_2018}. Under shear, molecular alignment and deformation can promote partially ordered regions and influence subsequent crystal growth.\cite{nie_precursor_2022,jariyavidyanont_thermal_2021,seki_shear_mediated_2002}
However, predicting how deformation strength and pre-shear duration affect crystallization remains a challenging task\cite{jabbarzadeh_flow-induced_2010}. Under deformation, polymer chains can undergo stretching and alignment \cite{sampath_extension}, which modify crystallization kinetics and influence the crystal structures that emerge \cite{yi2013molecular,mcilroy2017disentanglement,qu2016flow, hsu_standard_2010, mykhaylyk_specific_2008}. Flow can influence polymer crystallization not only by altering kinetics and final degree of crystallinity, but also by changing the morphology of the crystallites that form after deformation \cite{nie_recent_progress, graham_understanding_2019, wang2016flow, fawzi_morphology_2016,cryst7020051}. 

Although extensive theoretical \cite{graham_kmc, graham2019understanding,graham_modelling_2014} and experimental work has focused on either monodisperse melts or melts with narrow distribution\cite{vanMeerveld2004}, the role of molecular weight distribution remains less well understood, especially in bidisperse blends where short and long chains exhibit distinct relaxation and orientation dynamics under flow.\cite{guironnet_flow_chemistry, triandafilidi_molecular_2016,sampath_extension} Consequently, it remains difficult to predict how flow and dispersity together determine the resulting crystalline microstructure, including crystallite shape, organization, and connectivity. This highlights the need for simulations that can relate deformation history to the morphology of the resulting crystalline domains, as well as predictive models that connect processing conditions to microstructure \cite{ko_characterization_2004}.

Another relatively underexplored aspect is how flow history before crystallization affects the morphology of the crystalline domains that emerge, including their size, shape, and connectivity \cite{jiri_pre_shear,speranza_isothermal_shear_morphology}. In particular, it remains unclear how pre-shear influences not only the dimensions and anisotropy of crystallites, but also the molecular topology of the connections between them through bridging structures\cite{sharaf_lamella_connectivity,zhang_entanglement}. These features are expected to play an important role in cluster connectivity and subsequently stress transfer, yet their development under flow remains not well understood, especially in mixtures of short and long chains.

In this study, we investigate flow-induced crystallization in monodisperse and bidisperse polymer melts through segmentally coarse-grained molecular dynamics simulations of a model polymer. We provide a microscopic picture of how flow and molecular weight distribution shape crystalline morphology. The influence of pre-shear on crystallization kinetics and the final degree of crystallinity in melts is investigated and we analyze how chain length and composition affect size, shape, orientation, and connectivity of the resulting crystalline domains via bridging chains. Together, these results demonstrate how pre-shear and bidispersity influence crystalline organization, crystalline ordering, and molecular connectivity across domains in this simple coarse-grained model polymer.

The rest of this article is organized as follows. Section II presents the details of the segmentally coarse-grained model, resulting dynamics, rheological behavior of the chains, processing protocol, and the crystallinity identification algorithm. Section III discusses the results of the simulations and the shape, size, and connectivity characteristics of the chains. Finally, Section IV summarizes the main findings and provides an outlook.

\section{Methods}

\subsection{Model}

We performed molecular dynamics (MD) simulations to study polymer crystallization, conducting all simulations in the NVT ensemble using HOOMD-blue 5.1.0.\cite{anderson2020hoomd}  Homogeneous shear flow was enabled through a custom fork of HOOMD-blue implementing the SLLOD algorithm.\cite{todd2017nonequilibrium, edwards_validation_2006} Validation of this implementation, along with the corresponding source code, is provided in the Supporting Information (SI). All quantities are reported in a consistent system of reduced units with mass \(m\), length \(\ell\), and energy \(\varepsilon\); the time is then given by \(\tau=\sqrt{m\ell^{2}/\varepsilon}\). Each segment, or particle, had a mass of \(1.0\,m\).

Interactions between polymer segments, or particles, were modeled using a traditional pairwise DPD potential $U_{\text{pair}}$,\cite{dpd_hoomd_brownian, chertovich} with a harmonic bond potential describing chain connectivity,
\begin{align}
U_{\text{pair}}(r) &= A(r_{\text{cut}} - r) - \frac{A}{2}(r_{\text{cut}}^2 - r^2),\\
U_{\text{bond}}(r) &= k_{\text{bond}}(r - r_0)^2 \quad. 
\end{align}
The pair interaction strength was set at $A = 150\ \varepsilon/\ell$, while the bond potential employed a spring constant of $k_{\text{bond}} = 150\ \varepsilon/\ell^2$ and an equilibrium bond length of $r_0 = 0.2\ \ell$.The $r_{\text{cut}}$ was set to 1.0 $\ell$. 
This short bond length rendered the chains worm-like and satisfied a geometric no-crossing criterion suggested by Nikunen et al.\cite{nikunen,padding_uncrossability_2001},
 $ 2 r_{\min} > \ell_{\max}$,
where \( r_{\min} \) is the effective bead diameter.
Satisfying this inequality suggests topological uncrossability, leading to physically more realistic dynamics: short chains exhibit Rouse dynamics, whereas long chains become entangled and display reptation. \cite{nikunen,padding_uncrossability_2001} 

The simulation box contained particles at a number density of \(\rho_{\rm n} = 3/\ell^{3}\) and its length was set to $L = 30\ \ell$ in each direction. Initial polymer configurations were generated by a random-walk algorithm.\cite{sliozberg_equilibration} A displacement-capped method was used to remove initial strong overlaps and the system was subsequently equilibrated at \(k_{B}T = 1.5\,\varepsilon\). A shorter timestep of \(0.01\,\tau\) was used for all simulations to account for the stiffer bonds.

Temperature was controlled via a Nos\'e-Hoover thermostat with a coupling constant of \(1.0\,\tau\), ensuring uniform thermalization across the system even under pre-shear (see SI)\cite{yong2013thermostats}. Equilibration runs were performed at constant temperature \(T_\text{melt} = 1.5\,\varepsilon/k_{\rm B}\), and crystallization was induced by quenching the system to \(T_\text{cryst} = 1.0\,\varepsilon/k_{\rm B}\) which is below the equilibrium crystallization temperature of approximately \(T_c=1.15\,\varepsilon/k_{\rm B}\) (see phase diagram in Fig. S1 in the SI). Some systems were simulated with a pre-shearing step in addition to the temperature quench. The details are described in Section~\ref{sec:protocol}.

\subsection{Chain Dynamics and Melt Diffusion}

To identify the transition from the Rouse regime\cite{rubinstein2003polymer,kremer_dynamics_1990} to the reptation regime\cite{reptation}, mean squared displacements (MSD) were determined as
\begin{align}
\mathrm{MSD} = \frac{1}{5} \sum_{i=\frac{N}{2}-2}^{\frac{N}{2}+2} 
\left\langle \left[r_i(t) - r_i(0)\right]^2 \right\rangle.
\end{align}
Here, only the five innermost monomers were considered, as they better represent the mobility of the entire chain compared to the more mobile terminal monomers.\cite{kremer_dynamics_1990}
By comparing the resulting scaling behavior with the classical Rouse and reptation models,\cite{kremer_dynamics_1990,reptation, masubuchi2014simulating} we assessed the onset of entanglement in the simulated melts.\cite{zhang_entanglement,zou_zhang_entanglement,zhou_disentanglement_2013}

\begin{figure}[ht!]
     \includegraphics[width=\linewidth]{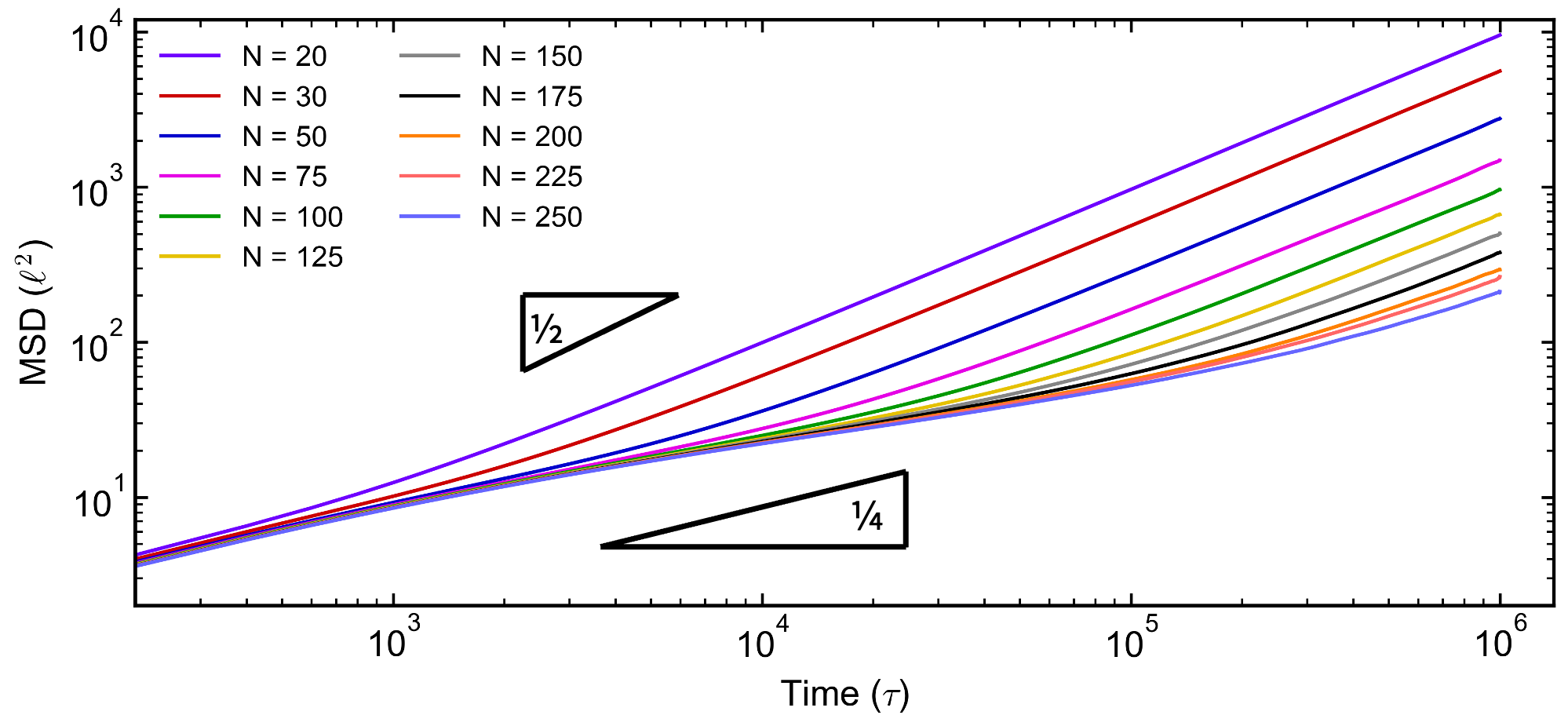} 
    \caption{Mean squared displacement of the five innermost segments for monodisperse polymer melts of various chain lengths $N$, at number density $\rho_{\mathrm{n}} = 3/\ell^{3}$ and temperature $T = 1.5\,\epsilon/k_{B}$.}
    \label{fig:msd2}
\end{figure}

As shown in Fig.~\ref{fig:msd2}, short chains exhibited a slope of $1/2$ in their MSD, characteristic of Rouse behavior,\cite{kremer_dynamics_1990} while longer chains with approximately $N \geq 125$ transitioned toward a $1/4$ scaling at intermediate times, indicative of reptation.\cite{reptation,nikunen}
This crossover reflects weak entanglement effects present in melts with soft pair interactions,\cite{iwata2002role,pan_developments_2003,padding_uncrossability_2001} consistent with other DPD-like coarse-grained models.\cite{holleran_2008, kumar_larson,faller_chain_2001, nikunen} Both diffusion coefficent and relaxation times exhibited scaling between the expected Rouse and reptation behavior, as shown in the SI.
The Flory characteristic ratio was found to plateau at $C_{\infty} \approx 4.2$ (see Fig. S2 in the SI) and the persistence length was $l_p \approx 1.1$–$1.2\,\ell$ (see Fig. S2 in the SI) independent of chain length, confirming that chains in our model exhibit a slight semiflexibility. These values are consistent with previously reported similar coarse-grained models of polyethylene-like polymers\cite{chertovich}.

\subsection{Processing protocol \label{sec:protocol}}

\begin{figure}[ht!]
    \includegraphics[width=\linewidth]{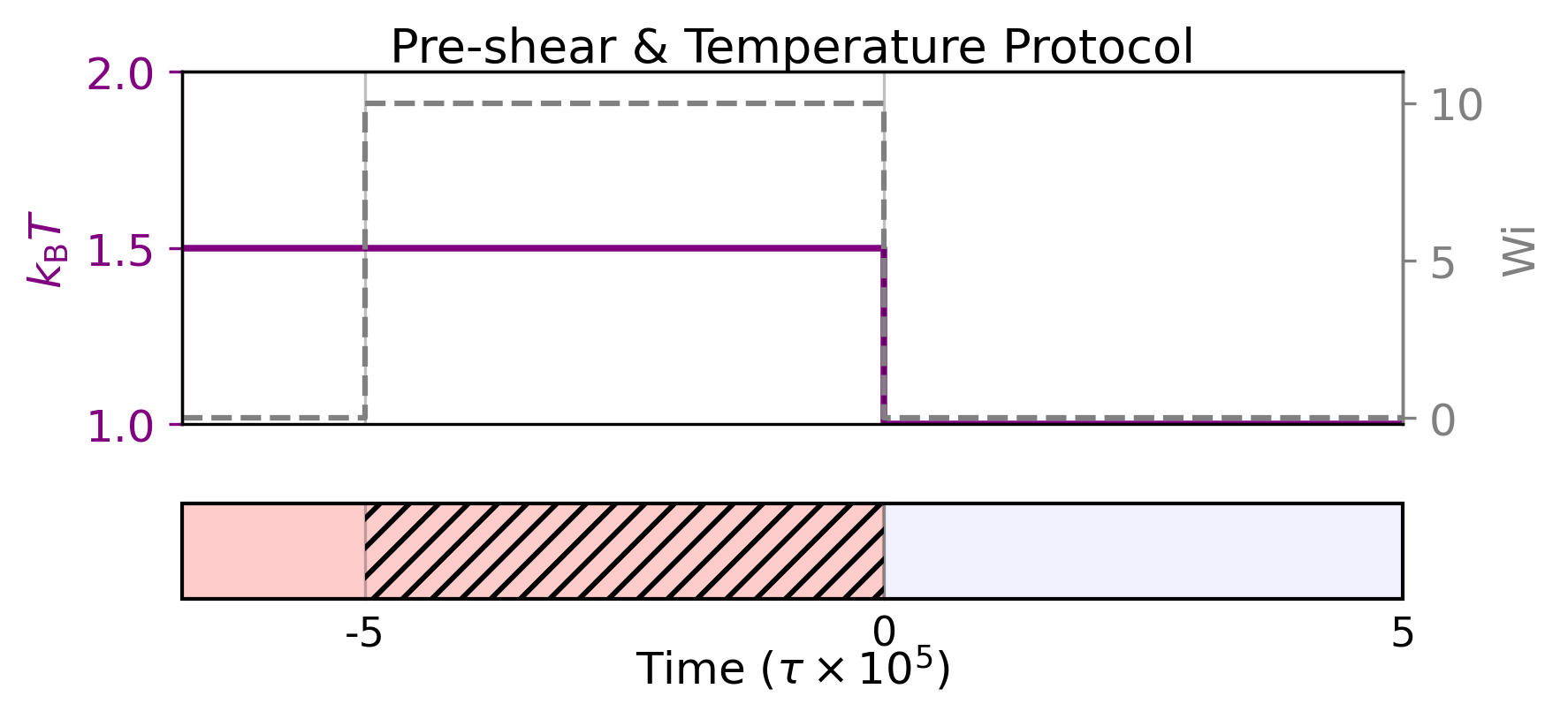}
    \caption{Shear and temperature protocol employed. First, equilibration for $10^6 \tau$ at $ 1.5\,\varepsilon/k_{\mathrm{B}}$ was performed. A pre-shear treatment followed for $5\cdot10^5 \tau$ at the same temperature, indicated by hatched lines in the colorbar and the dashed line in the figure. Finally, an instantaneous quench to $ 1.0\,\varepsilon/k_{\mathrm{B}}$ was performed at $t=0\ \tau$, after which the temperature was held constant for another $5\cdot10^5\tau$. }
    \label{fig:protocol}
\end{figure}

After equilibration, the systems were subjected to a controlled flow and temperature protocol designed to emulate typical polymer processing conditions. Both monodisperse (N = 30, 50, 100, 200) and bidisperse melts (N = 30-100, 50-100, 30-200, 50-200, chain pairs with equal weight fractions) were simulated. The number and weight averaged molecular weights of the bidisperse melts can be found in Table \ref{table:molecular_weights}. Each melt was equilibrated for $10^6 \tau$ at $T_\text{melt} = 1.5\,\varepsilon/k_{\mathrm{B}}$ to ensure fully relaxed intra-chain conformations (see Fig. S3 in the SI). This was followed by a pre-shear treatment production run of $5\cdot10^5 \tau$ at the same temperature, before quenching to $T_\text{c} = 1.0\,\varepsilon/k_{\mathrm{B}}$ for $5\cdot10^5\tau$. This temperature is below the crystallization temperature, \(T_c=1.15\,\varepsilon/k_{\rm B}\). The full protocol is depicted in Fig.~\ref{fig:protocol}. In the pre-shear step, flow was imposed in the box along the $x$-direction with the gradient along $y$, corresponding to a homogeneous SLLOD deformation. The intensity of flow was characterized by the Weissenberg number, 
\begin{equation}
\text{Wi} = \dot{\gamma}\,\tau_R \quad ,
\label{eq:weissenberg}
\end{equation}
where \( \dot{\gamma} \) 
is the applied shear rate and \( \tau_R \) is the relaxation time of the polymer chain. We first validated the quiescent behavior of our melts at $ \text{Wi}= 0$, i.e., at no pre-shear with a temperature quench only. For investigating the impact of pre-shear treatment, we then decided to simulate our melts under an applied flow sufficient to induce enhanced crystallization, at $\text{Wi} = 10$, where chain alignment increases significantly across the melt \cite{qu_flow-directed_2016}.

To determine the relaxation time, $\tau_R$, independent equilibrium simulations were performed for each melt composition. The mean-squared displacement (MSD) of chain centers of mass and the radii of gyration $R_g$ were measured to ensure that intrachain distances were fully relaxed and consistent with wormlike-chain statistics. 
The long-time diffusion coefficient $D$ was obtained from the slope of the MSD (see Fig. S4 in the SI) and the Rouse relaxation time was estimated as $\tau_R = R_g^{2}/D$ (see Fig. S4 in the SI). For bidisperse melts, the Weissenberg number was defined using the relaxation time of the long-chain component to ensure that chains with restricted mobility were sufficiently deformed by flow.

\begin{table}[h]
\centering
\begin{tabular}{c c c}
\textbf{Melt (short--long)} & $\mathbf{M_n} [m]$ & $\mathbf{M_w} [m]$ \\
\hline\hline
30--100 & 46.15 & 65 \\
\hline
50--100 & 66.67 & 75 \\
\hline
30--200 & 52.17 & 115 \\
\hline
50--200 & 80.00 & 125 \\
\end{tabular}
\caption{Number-average and weight-average molecular weights for the simulated bidisperse polymer melts assuming an equal mass split between short and long chains. Molecular weights are expressed in units of the bead mass $m$.}
\label{table:molecular_weights}
\end{table}

\subsection{Crystallinity identification algorithm}

Due to the soft, segmental nature of the model, crystalline regions in the simulations were generally less ordered than a model with hard-core repulsion would be. Nevertheless, we were able to identify crystal regions with a multi-step procedure. 

First, crystalline order was identified using a bond--orientational order criterion combined with a local density refinement. Crystalline particles were provisionally detected using Steinhardt bond--orientational\cite{lechner2008accurate} order parameters. Particle positions were unwrapped and a Voronoi tessellation was generated with the \texttt{freud}\cite{Voronoi1908,voro++,freud} library to obtain topology-aware nearest neighbors. Using these neighbors, we computed the Voronoi-weighted average Steinhardt order parameters \(\overline{q}_4\) and \(\overline{q}_6\).
Particles exceeding a common threshold \((\overline{q}_4,\,\overline{q}_6) \geq 0.25\) were provisionally labeled crystalline (see SI for histograms).\cite{q4_q6_threshold} 
To identify core crystalline regions within the subset of all crystalline particles, we also required that particles possess a sufficiently large number density of 1.0 crystalline neighbors within a sphere of radius \(2.0\ \ell\). Particles satisfying this condition were grouped into clusters using a distance cutoff of \(1.0\ \ell\).\cite{mickel_shortcomings_2013}

In a second step, these core crystal clusters were extended using alignment along the backbone. Intrachain bond vectors \((r_{i-1} \rightarrow r_{i+1})\) were used to define local backbone orientations of particle $i$, which were mapped to spherical angles \(\theta_i\) and \(\phi_i\). For each core cluster, mean values of \(\bar \theta\) and \(\bar \phi\) were computed. In a small number of iterative sweeps, neighboring crystalline beads were then added to the cluster if they were nearest neighbors of core crystalline regions and their backbone orientation deviated by less than \(10^\circ\) from the averaged values of that core crystalline region.

To further subdivide large crystallites into smaller orientationally coherent domains, i.e. crystal grains, we performed a third step \cite{vimal2021grainseg}. For each crystallite, all particles in that cluster  were unwrapped at the periodic boundary conditions. This ensured that only spatially continuous  clusters were considered for subsequent domain analysis.
Each particle $i$ within a crystallite was then represented by a 5D joint position--orientation feature vector,
$\mathbf{X}_i = (x_i, y_i, z_i, \phi_i, \theta_i)$,
combining spatial coordinates $(x_i,y_i,z_i)$ with backbone orientation $(\phi_i,\theta_i)$. Using these vectors, crystallites were then partitioned into subdomains using the density-based hierarchical clustering algorithm \texttt{HDBSCAN}.\cite{hdbscan} The clustering was performed with a minimum cluster size of 20 particles and a cluster selection tolerance 0.9 \(\varepsilon\), while allowing the identification of a single dominant cluster if appropriate. Particles labeled as noise by \texttt{HDBSCAN} were treated as unassigned or interfacial within the parent crystallite.

\begin{figure}[ht!]
    \includegraphics[width=\linewidth]{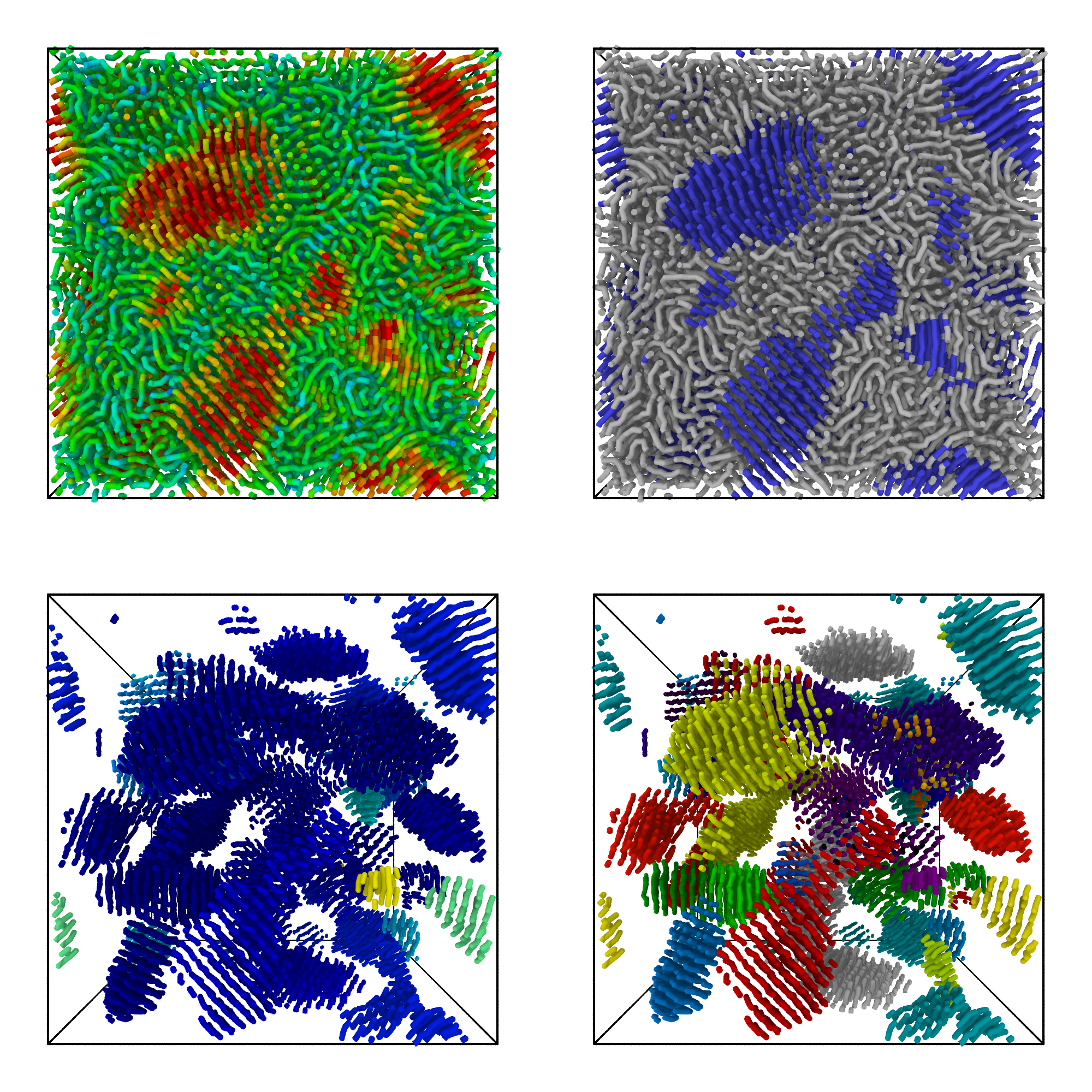}
    \caption{(Top left) Snapshot of the monodisperse melt with $N=50$ at $t=0.95\cdot 10^5 \tau$, where approximately 30\% of the system had crystallized. Color indicates the value of $\bar q_6$. (Top right) Same snapshot, colored by whether particles are identified as crystalline (blue) or amorphous (gray). (Bottom left) Same system but with amorphous sections removed for clarity. Colors indicate clusters found by a simple cutoff. (Bottom right) Same snapshot colored by grains identified using both location and orientation of particles and bonds.}
    \label{fig:crystal_grain_illustration}
\end{figure}

With this multi-step identification, we were able to determine the overall degree of crystallinity and individual grains, as shown in Fig.~\ref{fig:crystal_grain_illustration}. 
A simple distance-based clustering was not able to clearly distinguish individual crystal grains. In contrast, \texttt{HDBSCAN} operating on the combined position and orientation vector separated individual crystallite regions well, even in dense systems where different crystal grains are in direct contact. However, in this system with soft pair interactions significant noise was present. Therefore, further improvements in crystal grain identification would be beneficial.

\section{Results and Discussion}

\subsection{Crystallization Kinetics and Morphology}

\subsubsection{Overall crystallization in monodisperse and bidisperse melts}

\begin{figure}[ht!]
    \includegraphics[width=0.8\linewidth]{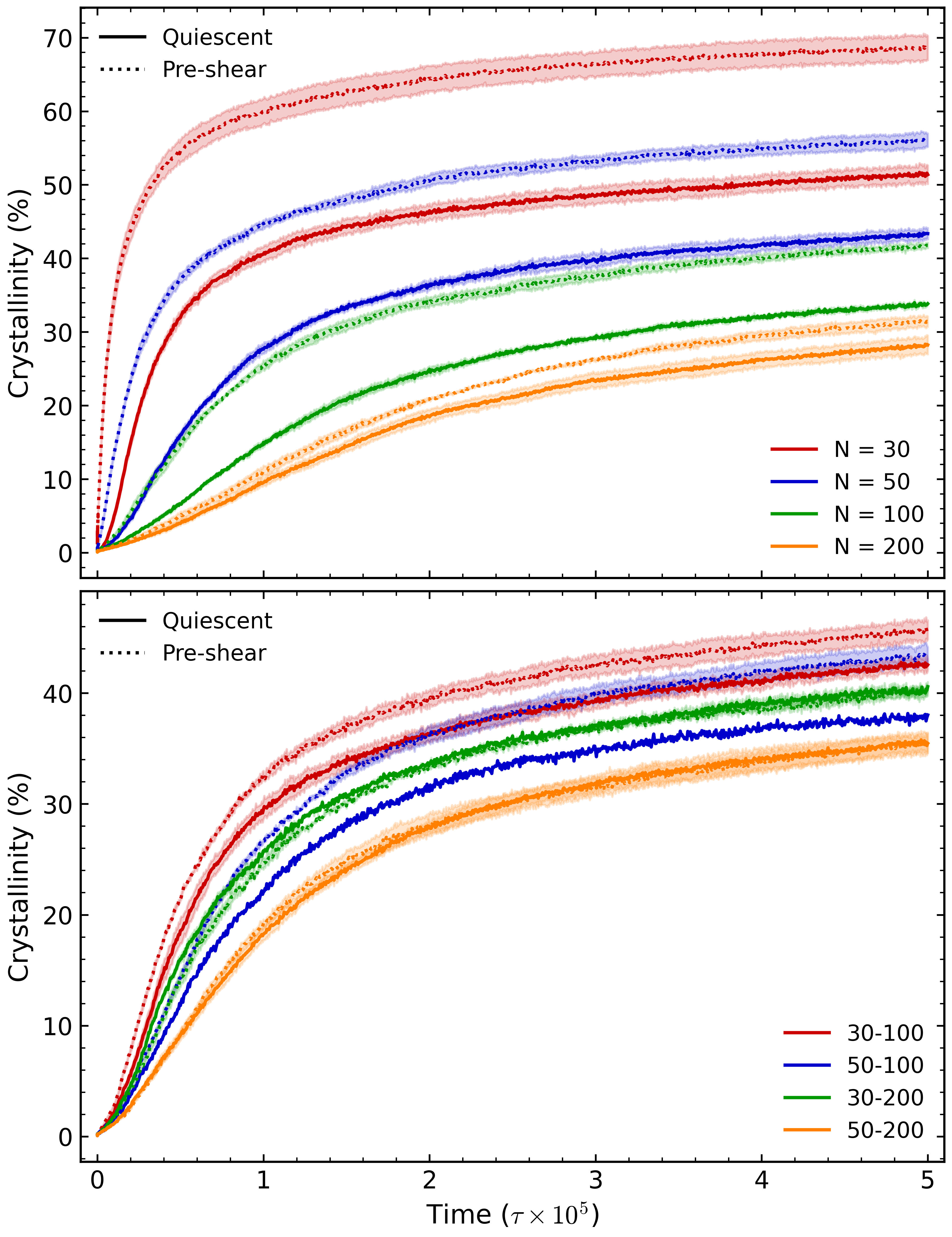}
    \caption{Time evolution of crystallinity for monodisperse melts (top) and bidisperse melts (bottom) under quiescent conditions (solid) and steady pre-shear at $\mathrm{Wi}=10$ (dotted). Shaded regions represent the standard error over three replicas.}
    \label{fig:crystal_mono_quiescent}
    \label{fig:crystal_bi_quiescent}
\end{figure}

Fig.~\ref{fig:crystal_mono_quiescent} shows the time evolution of the overall crystallinity after quenching from $T_\text{melt}=1.5\,\varepsilon/k_B$ to $T_\text{cryst}=1.0\,\varepsilon/k_B$ at $t=0\,\tau$ for monodisperse melts with $N=30, 50, 100,$ and $200$ on top, and bidisperse melts below. Both systems under quiescent conditions (solid lines) and systems with pre-shear at $\mathrm{Wi}=10$ (dotted lines) before quenching are shown. Data were averaged over three independent replicas; shaded bands indicate the standard error of the mean. In quiescent conditions, crystallinity increased monotonically over time and approached a chain-length-dependent plateau. Both the initial growth and the final value of crystallinity decreased systematically with increasing $N$, with final degrees of crystallinity roughly between 28\% (long chains) to 52\% (short chains). This trend is consistent with reduced chain mobility and increasing topological constraints for longer (slightly entangled) chains, which slow down structural rearrangements required for nucleation and growth\cite{sommer_molecular_2010}. 

Imposing pre-shear before quenching enhanced crystallization for all chain lengths by increasing initial growth of crystallinity, resulting in higher crystallinities, roughly between 32\% to 70\%. Notably, the magnitude of pre-shear induced crystallization diminished as chain length increased, even though, even though the relative strength of the pre-shear was adjusted for chain lengths by using a constant Weissenberg number $\text{Wi}=10$ for all systems. Short chains responded strongly to flow-induced orientation and packing during pre-shear, whereas longer chains exhibited modest to no gain, consistent with slower relaxation and increasing topological constraints that limit their ability to reorganize into crystalline order. Quantification of growth-rate differences and final values of crystallinity are reported below in the next section.

To probe the role of molecular weight dispersity, we simulated bidisperse blends consisting of short and long chains at equal mass fractions (Fig.~\ref{fig:crystal_bi_quiescent} bottom) under the same thermal and pre-shear protocol as the monodisperse systems. The Weissenberg number for bidisperse melts was calculated based on the relaxation time of the long chain component of the blend, using Eq. \ref{eq:weissenberg}. Under quiescent conditions, the bidisperse melts exhibited crystallization kinetics and plateaus that showed the same qualitative dependence on the presence of the long component; blends containing the longest component (200-mers) crystallized more slowly and reach lower crystallinity than blends where the long component was a 100-mer. However, the final values of crystallinity were all in the range of roughly 35\% to 43\%. 

In the case of bidisperse blends, crystallinity typically rose more quickly with pre-shear ($\mathrm{Wi}=10$) than in the quiescent case. However, this shear-induced enhancement was strongly composition-dependent: while melts with 100-mers showed a clear increase in crystallinity after pre-shear, the two melts containing 200-mers exhibited little measurable increase between pre-sheared and quiescent conditions. This weaker response is consistent with the idea that the long, highly constrained 200-mers remain kinetically trapped due to reduced mobility and entanglement constraints which limit their ability to align, diffuse, and incorporate into the growing crystalline domains, thereby suppressing any additional pre-shear induced crystallization\cite{luo_frozen_2014,zhang_entanglement,iwata2002role}.

\subsubsection{Growth Rate in Monodisperse and Bidisperse melts}

To quantify the trends observed in Fig.\ref{fig:crystal_mono_quiescent} further, we computed the time required for each system to reach 5\% crystallinity, denoted by $\tau_{5\%}$. Using this time, an early-stage growth rate proxy $k_{5\%}$ can be defined as its inverse, $k_{5\%} = 1/\tau_{5\%}$.
The $5\%$ threshold was chosen to characterize the onset of crystallization. Although the absolute values of the growth rate depend on the precise threshold selected, the qualitative trends and relative comparisons are expected to remain unchanged for reasonable choices of this threshold. 

\begin{figure}[ht!]
    \includegraphics[width=\linewidth]{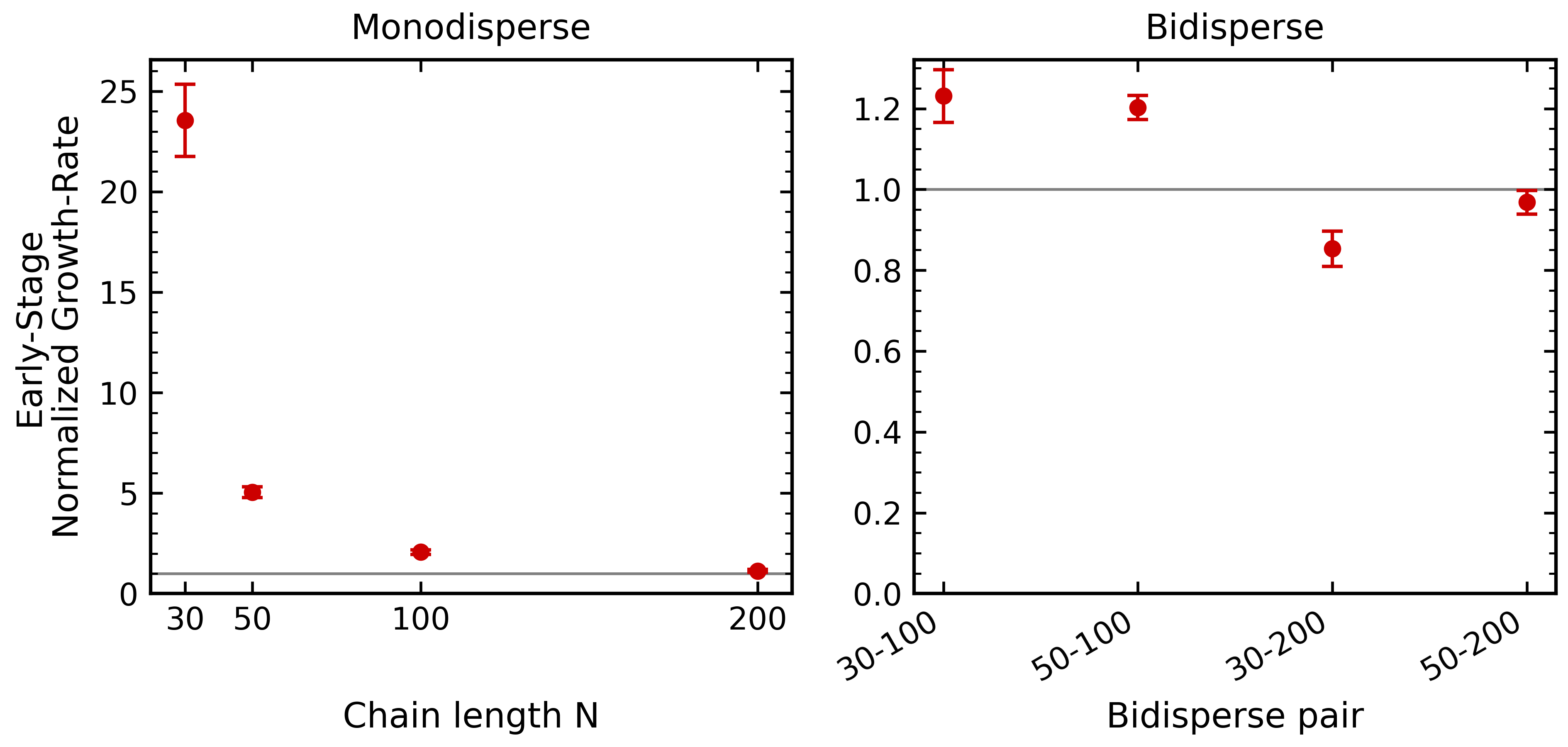}
    \caption{Relative rate of growth to achieve 5\% crystallinity in the quiescent system to the growth rate with pre-shear treatment. Values above unity indicate enhancement of crystalline growth with pre-shear. Results are shown for both monodisperse (left) and bidisperse systems (right) as function of chain length $N$. Error bars denote the standard error over three replicas.}
    \label{fig:normalized_rate_shear}
\end{figure}

In Fig.~\ref{fig:normalized_rate_shear}, the effect of pre-shear on the early-stage crystal growth rate is quantified through the normalized ratio $k_{5\%}(\mathrm{Wi}=10)/k_{5\%}(\mathrm{Wi}=0)$. Here, values above unity indicate that pre-shear enhanced the early-stage crystallization rate, whereas values below unity indicate a suppression of crystal growth relative to quiescent conditions. As shown in the left panel of Fig.~\ref{fig:normalized_rate_shear}, the monodisperse systems exhibited a clear enhancement for short chains and moderate or no enhancement for longer chains. In the bidisperse systems, a mild enhancement was observed for systems whose longest chains are $N=100$, whereas a mild suppression is found for systems containing 200-mers.

\begin{figure}[ht!]
    \includegraphics[width=\linewidth]{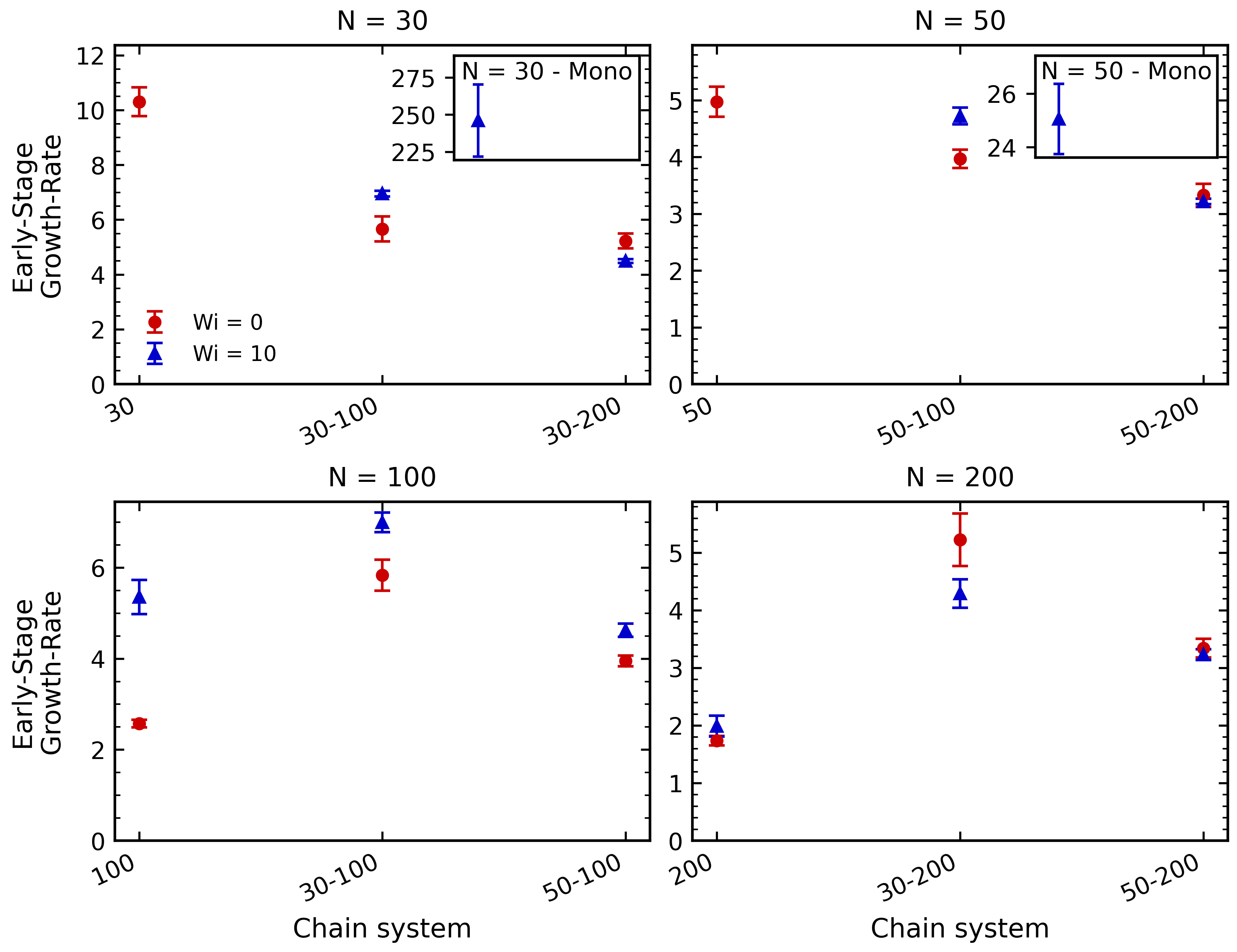}
    \caption{Component-resolved early-stage growth-rate  $k_{5\%}$ for selected chain populations in bidisperse melts, compared with the corresponding monodisperse melt (first point on the $x$-axis in each panel). For bidisperse systems, $k_{5\%}$ is determined from the relative crystallinity of the indicated component. Quiescent systems are shown in red and pre-sheared systems in blue. Each panel focuses on one specific polymer length, enabling direct comparison between that component in the monodisperse melt and in the corresponding bidisperse blends.}
    \label{fig:unnormalized_growth_relative_rates}
\end{figure}

To compare the effect of blending with shorter chains against the effect of pre-shear on crystallization kinetics, we evaluated the early-stage growth-rate proxy for individual chain populations in both monodisperse and bidisperse melts.
For monodisperse systems, $k_{5\%}$ was obtained directly from the total crystallinity of the melt. For bidisperse melts, we computed a component-resolved relative crystallinity for each chain population, defined as the fraction of crystalline particles belonging to a given component divided by the total number of particles of that same component in the blend. Thus, for a bidisperse melt containing short and long chains, the relative crystallinity of the short-chain component measures the fraction of short-chain particles that were crystalline, independent of the crystallization state of the long-chain component. The exact same definition was applied to the long-chain component. The corresponding component-resolved growth-rates were then extracted from the time required for each component to independently reach 5\% relative crystallinity.\cite{triandafilidi_molecular_2016}

From these results, shown in Fig.~\ref{fig:unnormalized_growth_relative_rates}, one can conclude that adding short chains to a melt had a greater effect on long chain crystal growth rates than pre-shear.
In contrast, for the short-chain components, the monodisperse melts exhibited substantially larger growth-rates than the corresponding bidisperse systems.

\subsubsection{Final Crystallinity and Cluster Size }

While our simulations are finite in time due to computational limitations, extracting the final values still 
provides a compact view of how pre-shear and molecular size set the final crystallinity and the typical size of crystalline aggregates, or grains. For this, we measured properties at the end of the simulation, after $5\cdot 10^5 \tau$. From Fig. \ref{fig:crystal_bi_quiescent} it is clear that there could be additional slow crystal growth that might be observed if the simulations had been run for even longer. However, trends and relative ordering of the systems are not expected to change significantly. 

\begin{figure}[ht!]
    \includegraphics[width=\linewidth]{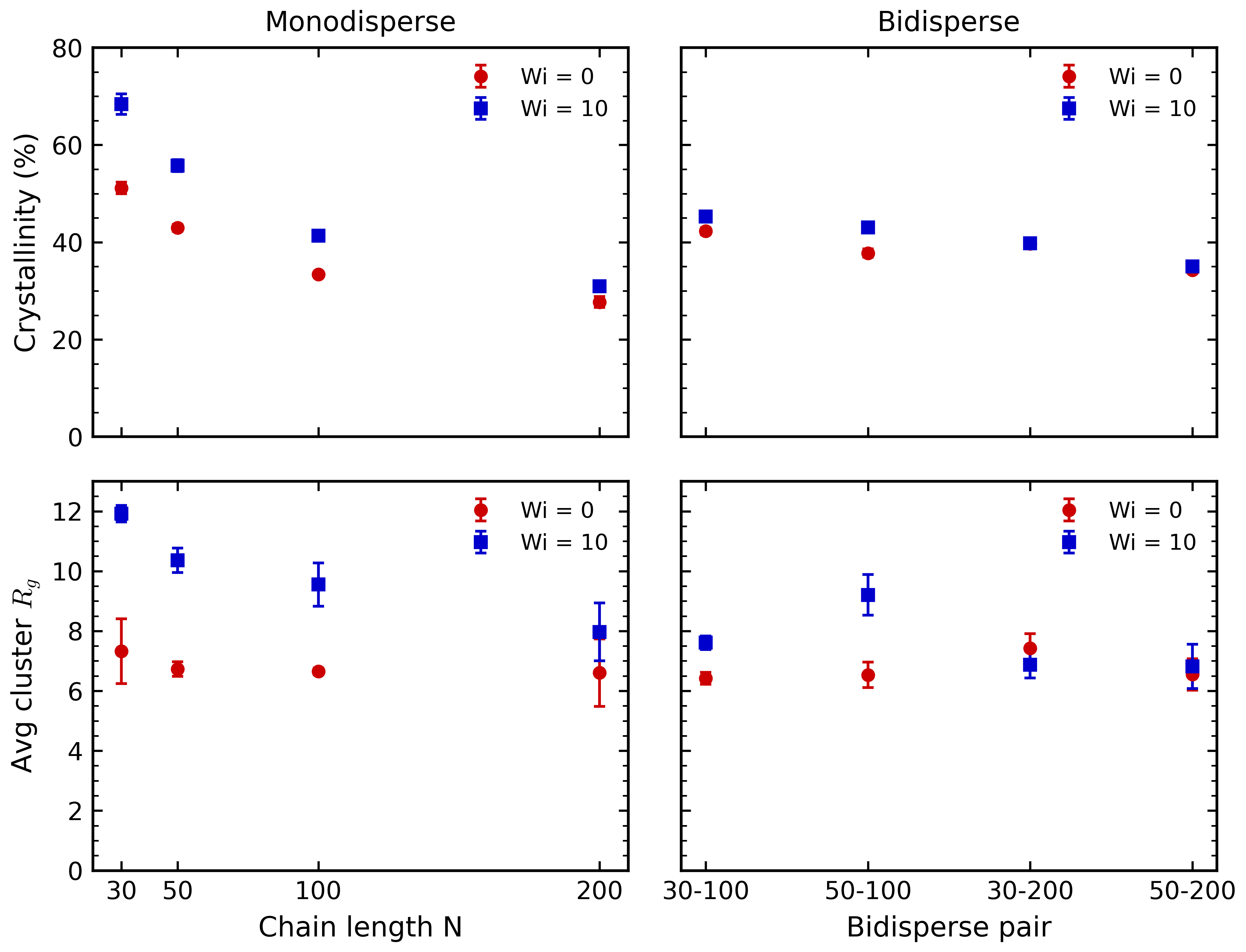}
    \caption{Final values of crystallinity and average grain size in monodisperse and bidisperse melts under quiescent conditions (Wi = 0) and pre-shear (Wi = 10).}
    \label{fig:end_values_crystallinity_cluster_no}
\end{figure}

For monodisperse melts, the final crystallinity (top panels of Fig. \ref{fig:end_values_crystallinity_cluster_no}) decreased systematically with increasing chain length, and pre-shear yielded consistently higher terminal crystallinity than quiescent conditions, with the largest enhancement observed for the shorter chains. The same trend was reflected in the final average grain size (bottom panels of Fig. \ref{fig:end_values_crystallinity_cluster_no}), which was clearly increased under shear for $N = 30$ and $N=50$, indicating the formation of larger terminal aggregates. In contrast, the effect of flow on grain sizes was much weaker for the 100-mers and the 200-mers, consistent with a reduced ability of pre-shear to promote grains merging at longer chain lengths. It is also interesting to note that under quiescent conditions, the chain length and blend dispersity had little to no effect on the overall grain size.

For bidisperse melts, the final crystallinity values showed only modest dependence on mixture composition and the effect of flow on the terminal crystallinity remained modest across all chain blends (Fig. \ref{fig:end_values_crystallinity_cluster_no}). The final average grain size in bidisperse systems was also less responsive to flow than in the short-chain monodisperse melts and was basically unchanged when the mixture includes the longest component (blends containing 200-mers). Taken together, these summary plots reinforce the kinetic picture from the time-resolved analysis: pre-shear most effectively increased the final grain sizes in short-chain monodisperse melts, whereas bidispersity, particularly in mixtures containing very long chains, limited the extent to which flow can restructure the final grain sizes and overall crystallinity.

\begin{figure}[ht!]
    \includegraphics[width=\linewidth]{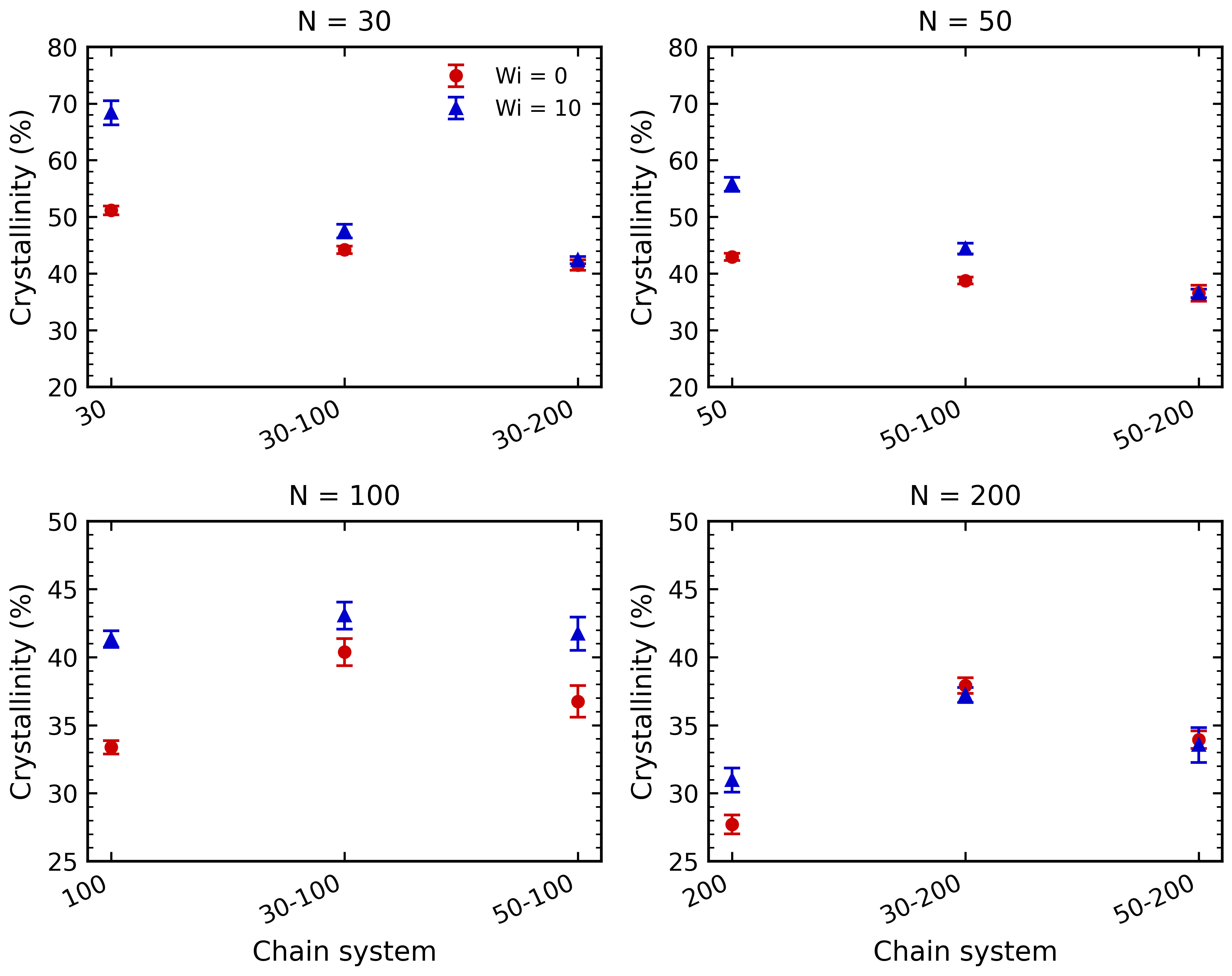}
    \caption{Component-wise relative final crystallinity in the bidisperse melts. Each panel focuses on one chain length, showing both quiescent (red) and pre-sheared (blue) systems.}
    \label{fig:end_values_crystallinity_cluster_comparison}
\end{figure}

Interestingly, although the component-resolved growth-rate proxy $k_{5\%}$ was clearly affected by composition and pre-shear, the corresponding final relative crystallinities varied less strongly across the bidisperse blends. Here, the relative crystallinity of a given component was defined as the number of crystalline particles belonging to that component divided by the total number of particles of the same component in the melt \cite{triandafilidi_molecular_2016}. Thus, in each panel of Fig.~\ref{fig:end_values_crystallinity_cluster_comparison}, the monodisperse points correspond to the total crystallinity of the monodisperse melt, while the bidisperse points report the component-resolved relative crystallinity of the indicated chain population within the blend. For the short-chain components ($N=30$ and $N=50$), the monodisperse melts exhibited the highest final crystallinity, particularly under pre-shear, whereas the corresponding short-chain relative crystallinities in bidisperse melts were lower and only weakly affected by pre-shear. For the long-chain components ($N=100$ and $N=200$), blending with shorter chains led to higher final relative crystallinity than in the corresponding monodisperse melts, while pre-shear produced only a modest additional increase or, in some cases, no effect. Overall, the trends in final relative crystallinity were qualitatively consistent with those observed for the component-resolved growth-rate proxy, but the differences were less pronounced. Under the present conditions, adding short chains, particularly $N=30$, had a stronger effect on the final crystallinity of the long-chain component than pre-shear alone. This is most clearly observed for the $N=200$ chains.

\subsection{Cluster Topology and Chain Connectivity}
\subsubsection{Ties \& Loops}

\begin{figure}[ht!]
    \includegraphics[width=0.6\linewidth]{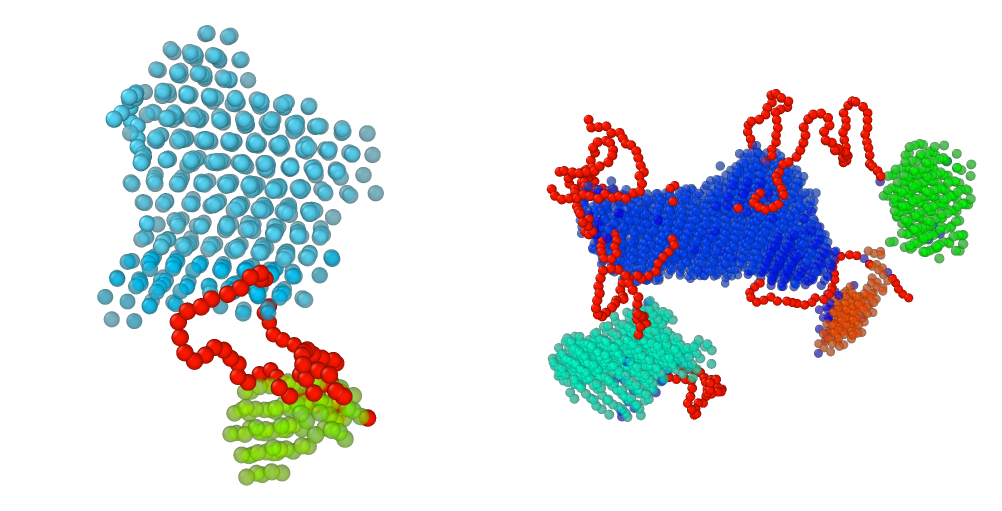}
    \includegraphics[width=0.5\linewidth]{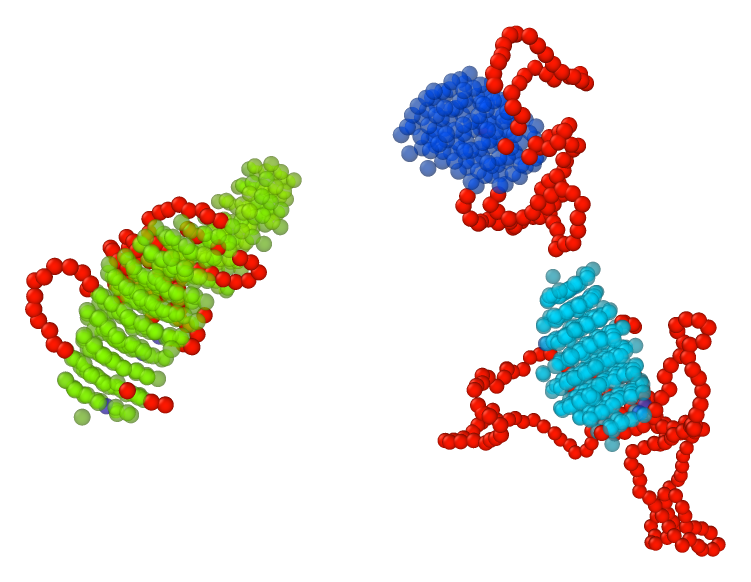}
    \caption{(Top) Tie chain formation connecting different crystal grains. Amorphous parts of the tie chains are highlighted with red, depict bridging of multiple crystallites through these chains, that act as stress transmitters in semi-crystalline polymers. (Bottom) loop chains are highlighted in red.}
    \label{fig:monodisperse_topology}
\end{figure}

\begin{figure}[ht!]
    \includegraphics[width=\linewidth]{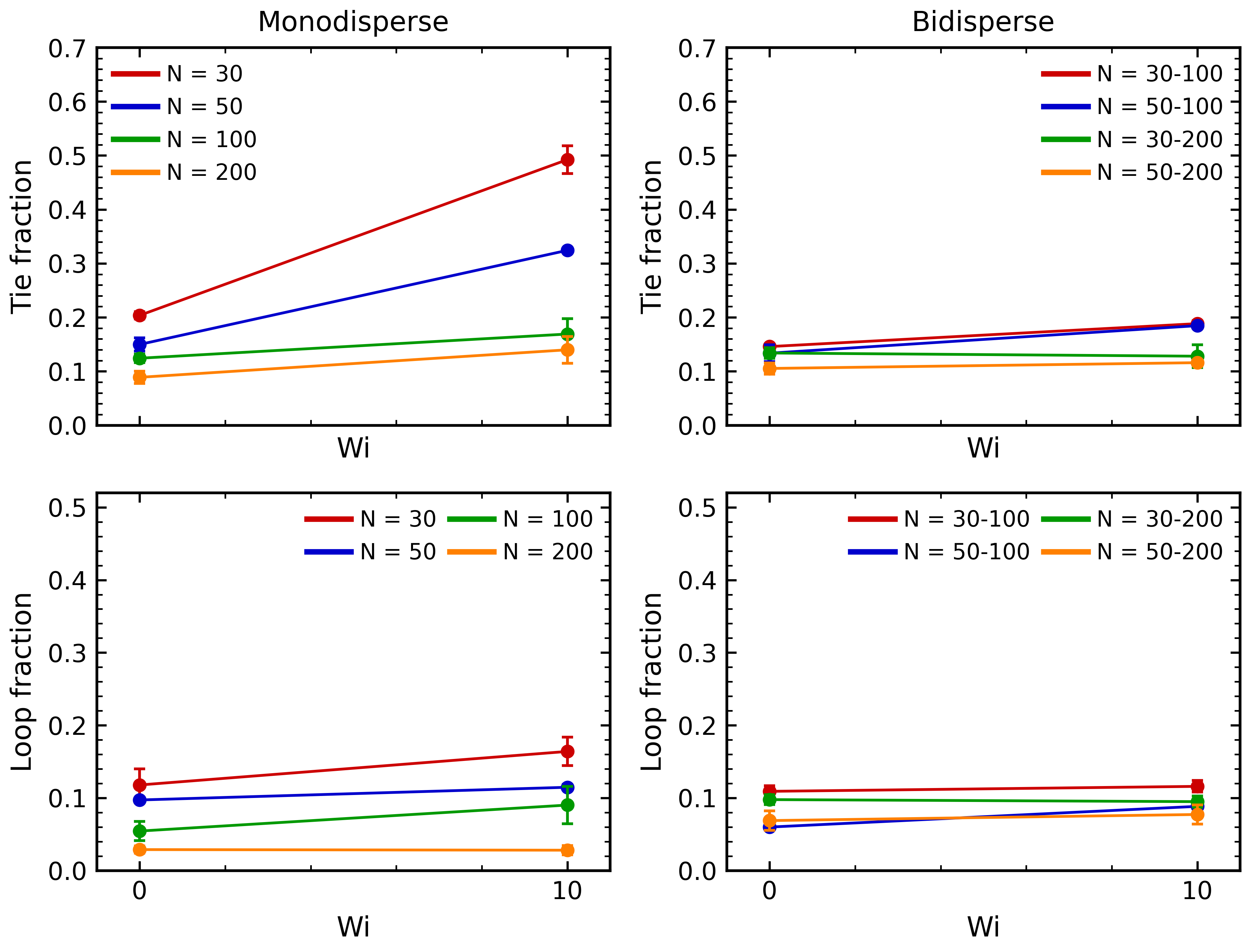}
    \caption{Fraction of tie and loop chains over total number of chains in monodisperse and bidisperse melts with varying Weissenberg number.}
    \label{fig:tie_loop_Wi}
\end{figure}

To connect crystallization kinetics to chain-level connectivity, crystalline chains were classified according to how they link crystalline grains. Loop chains have both ends incorporated within the \emph{same} crystalline grain with an amorphous segment in between, whereas tie chains connect two \emph{distinct} crystalline grains through an intermediate amorphous portion. Examples of both tie (top) and loop (bottom) chains are shown in Fig.\ref{fig:monodisperse_topology}. These populations provide a measure of whether crystallization primarily promotes intra-domain chain folding or inter-domain chain bridging.\cite{zhang_entanglement} Fig.\ref{fig:tie_loop_Wi} summarizes the dependence of tie and loop populations on Weissenberg number for both monodisperse and bidisperse systems. The same data normalized by crystallinity (Fig. S9 in the SI) showed that the trends were not an effect of increased crystallinty fraction. The time evolution of tie and loop fraction in monodisperse and bidisperse blends can be found in Fig. S10 and S11 in the SI.

For monodisperse melts, the overall tie chain fraction increased significantly with pre-shear treatment ($\mathrm{Wi}=10$) compared to quiescent conditions (Fig. \ref{fig:tie_loop_Wi}). This enhancement was most pronounced for the shorter chains ($N=30$ and $N=50$), where the tie fraction nearly doubled, while longer chains ($N=100$ and $N=200$) exhibited a more moderate increase. These results indicate that shear promotes the formation of inter-domain bridges, particularly in systems composed of shorter chains.

In contrast, the loop-chain fraction exhibited a weaker sensitivity to shear. For monodisperse melts, loop populations increased only modestly with $\mathrm{Wi}$, and the relative change was substantially smaller than that observed for tie chains. This distinction suggests that flow primarily enhanced connectivity between distinct crystalline grains rather than simply increasing intra-domain folding.
Additionally, the loop fractions for all systems were quite low. We attribute this to the worm-like character of the segmentally coarse-grained polymer model used in this study. For an increased ability to fold back on itself, a coarse-grained chain model with distinct angle potentials would be needed.~\cite{pva_muller_plathe,pva_zhao}

Generally, bidisperse mixtures showed a weaker effect of pre-shear treatment for both loop and tie fractions, mirroring the overall trends observed in the other properties. 
The blends displayed a small composition-dependent response to pre-shear. Blends containing the shorter $N=30$ chains showed a minor increase in tie chain fraction under shear, while mixtures involving longer components (30--200 and 50--200) exhibited almost no change. Loop chain fractions in bidisperse systems also showed only minor variations with $\mathrm{Wi}$. 

\subsubsection{Shape Characteristics}

The shape of crystalline grains was quantified using their prolateness\cite{singh_prolateness}, $P$ and asphericity normalized by the radius of gyration, $AS/R_g^2$.\cite{venetsanos_shape} Because the radius of gyration squared \(R_g^2\) measures the overall grain size, it can be used for normalization. $AS/R_g^2$ then quantifies the degree to which the cluster is \textit{spherically} symmetric, i.e., $AS$ is only zero for perfectly symmetric clusters like spheres, cubes, or tetrahedrons. Deviations from zero indicate non spherically distributed particles, with larger values indicating increasingly elongated or anisotropic clusters. The prolateness \(P\) distinguishes the nature of this anisotropy: positive values correspond to prolate, rod-like clusters, whereas negative values indicate oblate, disk-like clusters. Values close to zero correspond to more nearly isotropic shapes.

\begin{figure}[ht!]
    \includegraphics[width=0.9\linewidth]{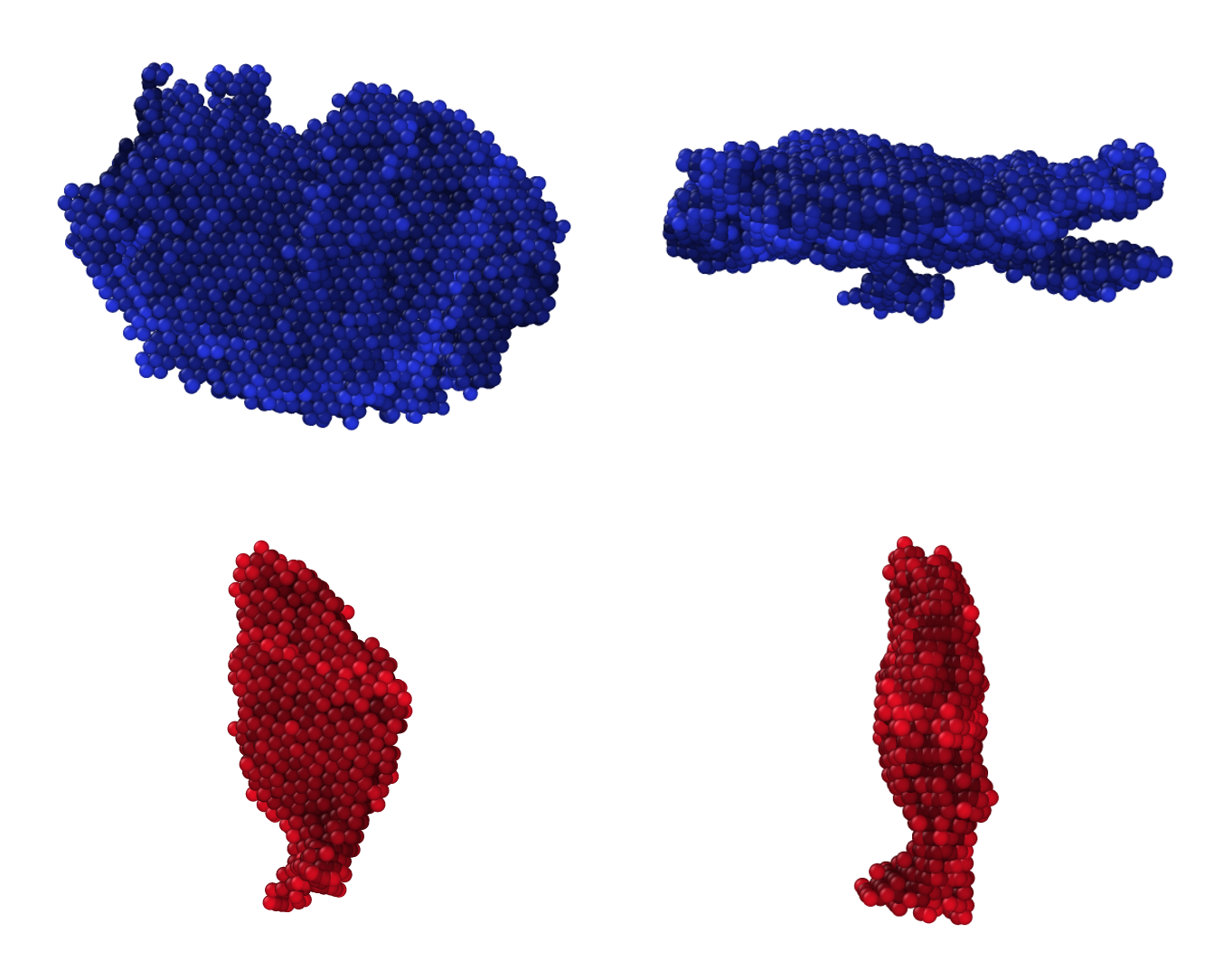}
    \caption{Comparison of cluster shapes of monodisperse systems formed under pre-shear (top, blue) and quiescent (bottom, red) conditions. The pre-sheared system produces an oblate, disk-like morphology with reduced thickness, whereas the quiescent system yields a more rod-like structure.}
    \label{fig:disk}
\end{figure}

\begin{figure}[ht!]
    \includegraphics[width=\linewidth]{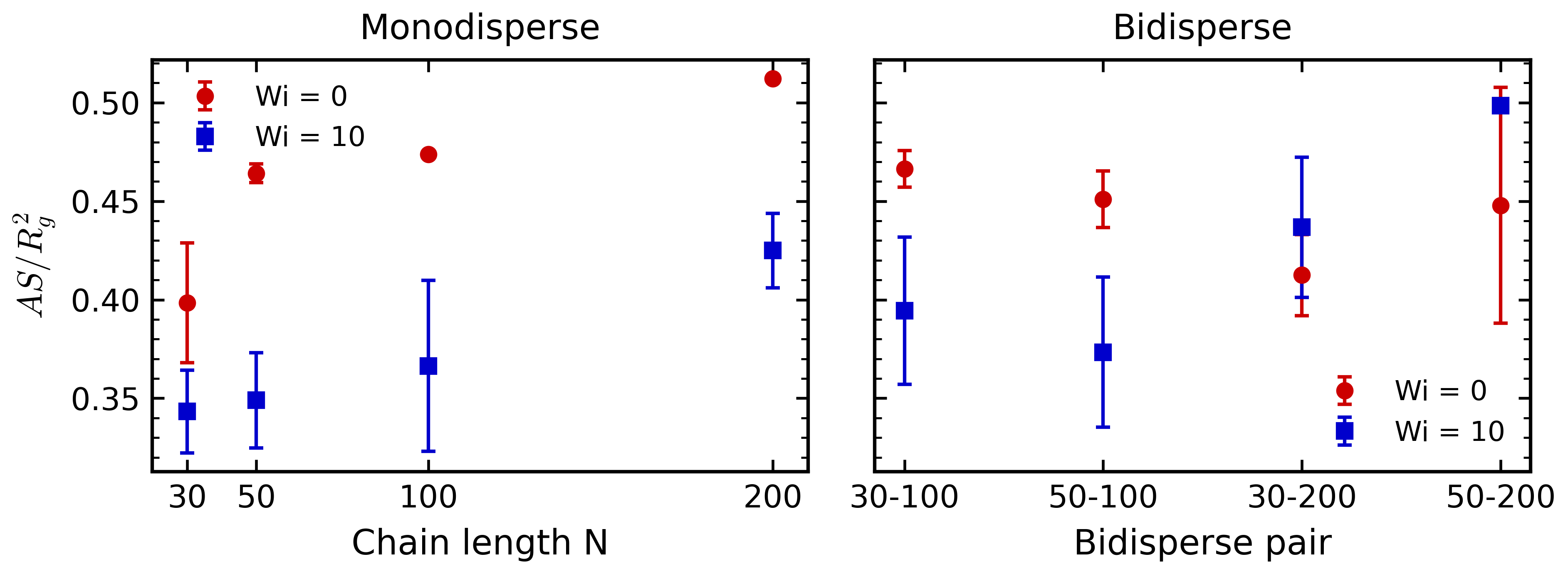}
    \includegraphics[width=\linewidth]{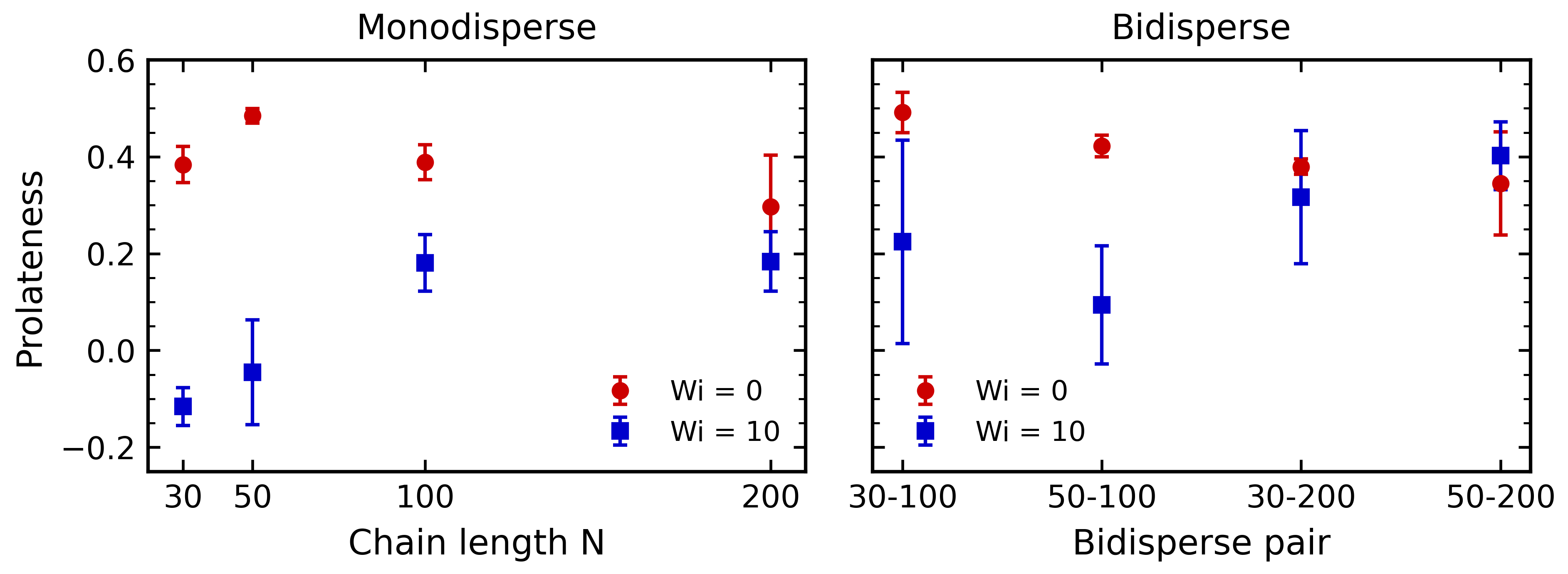}
    \caption{(Top) Acylindricity $AS/R_g^2$ of the clusters formed in bidisperse melts at quiescent conditions (Wi = 0) or under an imposed shear flow field (Wi = 10). (Bottom) Prolateness $P$ of the clusters formed in monodisperse  \& bidisperse melts at quiescent conditions or under an imposed shear flow field.}
    \label{fig:bidisperse_topology}
\end{figure}

The behavior of $AS/R_g^2$, taken together with the prolateness $P$ showed a clear trend for the monodisperse melts. For $N=30$, $50$, and $100$, pre-shear treatment reduced $AS/R_g^2$ relative to $\mathrm{Wi}=0$, indicating that clusters became more geometrically isotropic when pre-shear was applied (see Fig. \ref{fig:bidisperse_topology}). Visual inspection (see Fig. \ref{fig:disk}) confirmed that monodisperse melts under shear were more disk-like and were often extending over the periodic boundary conditions in one or two dimensions. 
The short dimension was commonly oriented along the backbone of the polymers, as displayed by the snapshots of representative clusters in Fig.~\ref{fig:disk}. As shown in Fig. \ref{fig:end_values_crystallinity_cluster_no}, clusters under shear were generally larger, in agreement with this observation. This visual assessment is also supported by a negative prolateness value for the shortest two chains with pre-shear.
In contrast, without pre-shear,  rod-like clusters were formed as indicated by $P\approx 0.4$, a representative example is shown in Fig.~\ref{fig:disk}.
For the longest chains ($N=200$), the difference between $\mathrm{Wi}=0$ and $\mathrm{Wi}=10$ was considerably smaller than for shorter chains.

For bidisperse blends, pre-shear had some impact on cluster shape. Overall, all clusters were generally rod-like, indicated by a positive prolatenes around 0.2 to 0.4 and an $AS/R_g^2$ around 0.35 to 0.45. Nevertheless, the same trend as in the monodisperse system is visible: the two systems with $N=100$ had lower prolateness values under pre-shear treatment, although they did not reach negative values.
For the blends containing the longest chains (30-200 and 50-200), no significant difference in shape parameters were found. 

$AS/R_g^2$ decreased under shear for mixtures containing shorter chains (30--100 and 50--100). In mixtures containing the longest chains (30--200 and 50--200), the pre-shear induced change in $AS/R_g^2$ was comparatively weak, indicating a reduced sensitivity of cluster shape to flow in these  blends with longer chain lengths.

\section{Conclusion}

In this study, we employed a soft, segmentally coarse-grained polymer model to investigate the effect of pre-shear on crystallizing monodisperse and bidisperse polymer melts. By systematically varying chain length, molecular weight distribution, and applying pre-shear, we determined how molecular weight distribution and flow can affect the overall crystallinity, relative crystallinity of each component, growth rates, grain size, and grain shape in melts and blends. We also investigated the effect on crystalline topology and chain alignment during crystallization.

Across the systems studied, bidispersity had a stronger influence on crystallization kinetics and final crystallinity than pre‑shear treatment; adding short chains to a melt of longer chains increased the final degree of crystallinity by approximately 10\% and accelerated the initial growth rate by roughly a factor of two. In contrast, pre‑shear of the melt prior to quenching led only to modest increases in growth rates and crystallinity, with a notable exception for monodisperse melts of short chains. Pre‑shear did, however, systematically affect crystalline morphology by reducing grain asphericity and prolateness in most monodisperse and bidisperse systems, instead forming more disk-like, isotropic crystallite shapes with pre-shear.

Beyond bulk crystallinity and grain shape, our analysis of molecular connectivity highlighted how pre-shear and dispersity influence the topology of crystalline domains. Pre‑shear increased the fraction of tie chains significantly in monodisperse systems. Bidisperse melts showed comparatively small changes in chain populations despite clear differences in crystallization kinetics and morphology. These results suggest that within this model, molecular weight distribution plays an important role in determining both crystallization dynamics and network connectivity.
 
Despite providing detailed microscopic insight into flow-induced crystallization in mono- and bidisperse polymer melts, several limitations are inherent to the modeling approach. First, the polymer model employs soft pair interactions, meaning that precise atomistic structure and energy dissipation mechanisms present in atomistic systems are missing. While this approach has been shown to reproduce qualitative entanglement and crystallization behavior in coarse-grained melts~\cite{chertovich}, quantitative crystallization kinetics depend on the choice of model. Second, the melts investigated here are weakly entangled, such that reptation dynamics emerge only for the longest chains studied. As a result, the interplay between strong entanglement and extreme molecular-weight disparities in blends  require further investigation. In addition, we only focused on homogeneous shear flow under isothermal quench conditions; more complex deformation histories (e.g., extensional flow) are likely to induce distinct crystallization pathways and morphologies. Finally, finite simulation lengths and finite system sizes limited access to very long-range lamellar organization and late-stage coarsening. 

Together, our findings offer a microscopic view of how flow history and molecular architecture interact to change crystallization in semicrystalline polymers. They provide molecular insight into how dispersity can be tuned to modulate crystallization kinetics and structural anisotropy. The methodology developed for resolving individual crystal grains and chain topology determination could also be expanded to probe the mechanical implications of tie and loop chain networks, offering a potential direct link between processing, microstructure, and macroscopic response.

\section{Acknowledgments}
This work was supported by the donors of ACS Petroleum Research Fund under Doctoral New Investigator Grant 65334-DNI7. AS served as Principal Investigator on ACS PRF 65334-DNI7 that provided support for TK.
This work made use of the Illinois Campus Cluster, a computing resource that is operated by the Illinois Campus Cluster Program (ICCP) in conjunction with the National Center for Supercomputing Applications (NCSA) and which is supported by funds from the University of Illinois at Urbana-Champaign.
This work also used Delta at the National Center for Supercomputing Applications through allocation CHM250093 from the Advanced Cyberinfrastructure Coordination Ecosystem: Services \& Support (ACCESS) program \cite{access}, which is supported by National Science Foundation grants.

\bibliography{references}

\pagebreak
\newpage
\onecolumngrid

\setcounter{equation}{0}
\setcounter{figure}{0}
\setcounter{table}{0}
\setcounter{page}{1}
\makeatletter
\renewcommand{\theequation}{S\arabic{equation}}
\renewcommand{\thefigure}{S\arabic{figure}}

\begin{center}
\textbf{\large Supplemental Information: Effect of Pre-Shear and Dispersity on Crystallization of a Model Polymer with Soft Pair Interactions using Molecular Dynamics Simulations}
\end{center}

\section{Polymer melt properties}

\begin{figure}[!htbp]
    \centering
    \includegraphics[width = 0.5\textwidth]{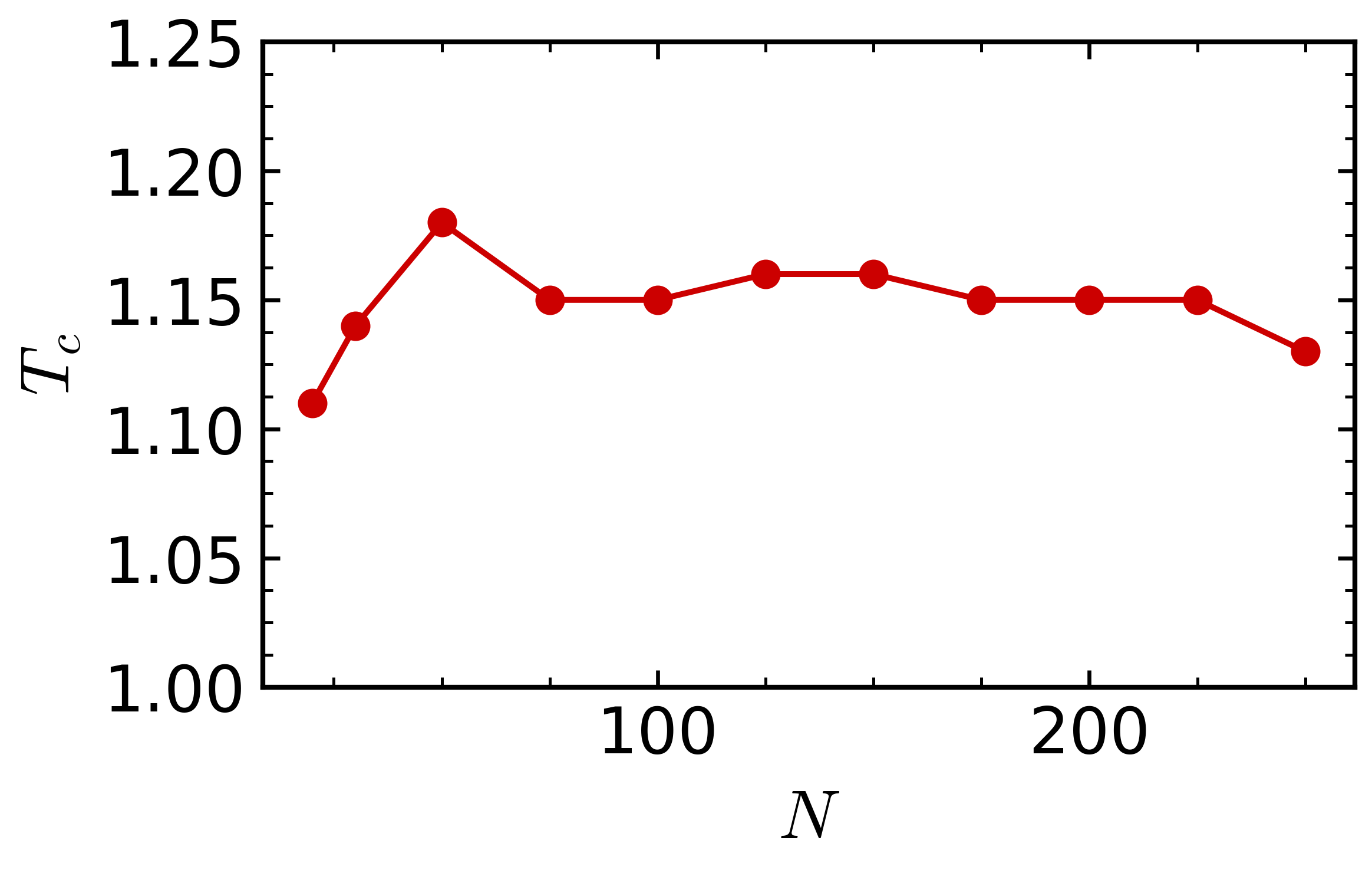} 
    \caption{Crystallization temperature $T_c$ as function of chain length $N$.}
    \label{fig:SI_phase_diagram}
\end{figure}

We determined the critical temperature of the melts studied here by slowly cooling them at a rate of $1.25\times 10^{-5}\,\varepsilon/(k_{\rm B}\tau)$. The crystallization temperature $T_c$ was determined by using a threshold of either 1\% overall crystallinity in the melt, which resulted in a chain-length independent estimate of $T_c \approx 1.15 \,\varepsilon/k_B$, as shown in Fig.~\ref{fig:SI_phase_diagram}. 

To quantify chain stiffness, we extracted two complementary measures from the polymer trajectories: Flory's characteristic ratio, $C_{\infty}$, and the persistence length, $L_p$. Flory's characteristic ratio was obtained from the average cosine of the angle between successive bond vectors along the chain backbone. Using the freely rotating chain relation,
\[
C_{\infty} = \frac{1 + \langle \cos \theta \rangle}{1 - \langle \cos \theta \rangle},
\]
this metric captures the local stiffness imposed by nearest-neighbor bond orientations. In contrast, the persistence length was determined from the decay of orientational correlations along the chain contour. Specifically, we computed the tangent-tangent correlation function, $\langle \mathbf{u}(0)\cdot\mathbf{u}(s)\rangle$, as a function of contour distance and extracted $L_p$ from the exponential decay of this correlation at short separations, using \texttt{freud}\cite{freud},
\[
\langle \mathbf{u}(0)\cdot\mathbf{u}(s)\rangle \sim e^{-r/L_p}.
\]
Together, these two quantities provide consistent but distinct descriptions of chain stiffness: $C_{\infty}$ characterizes local bond-angle correlations, while $L_p$ measures the distance over which backbone orientation remains correlated.

\begin{figure}[!htbp]
    \centering
    \includegraphics[width=0.4\textwidth]{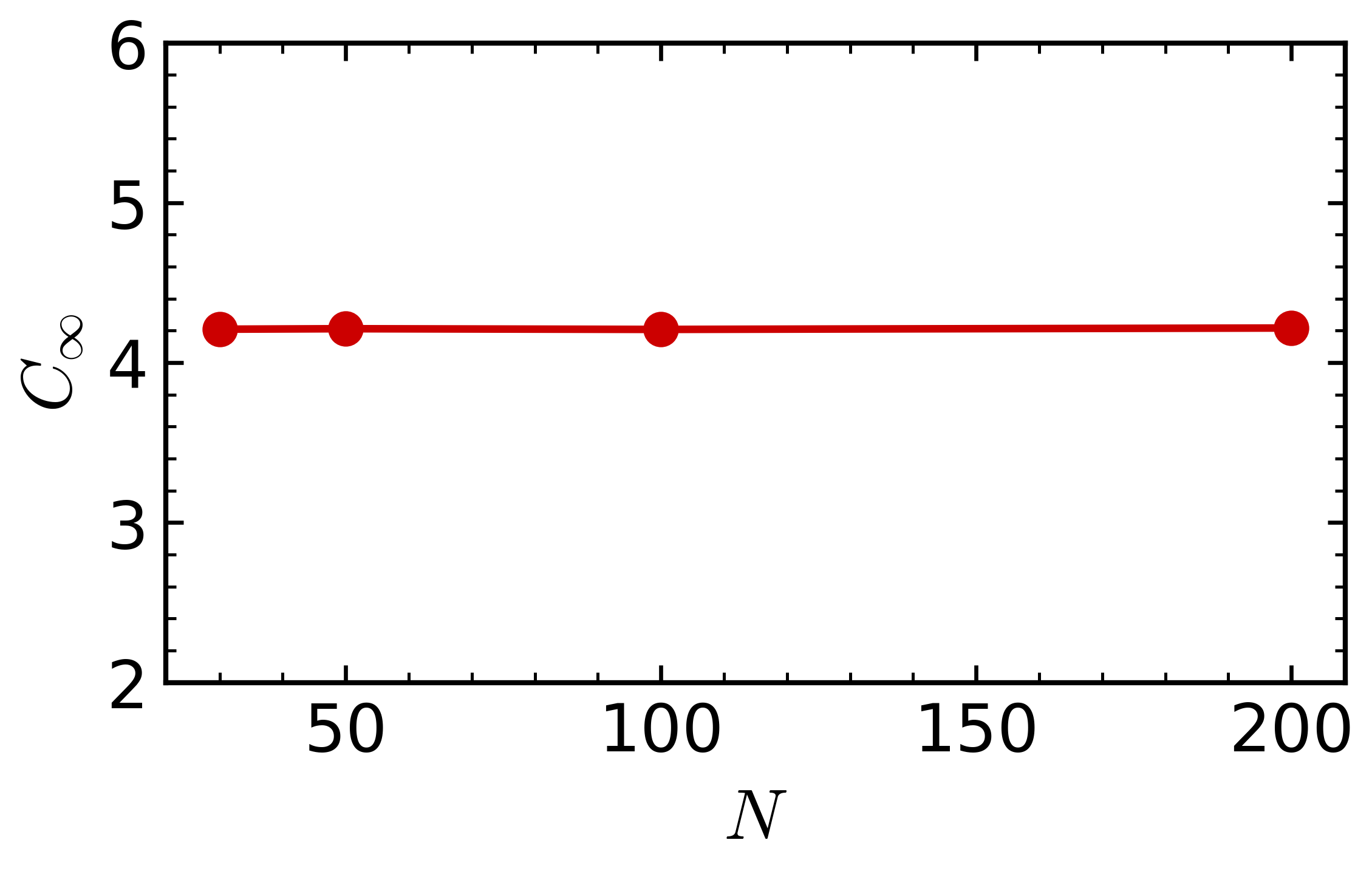}
\includegraphics[width=0.4\textwidth]{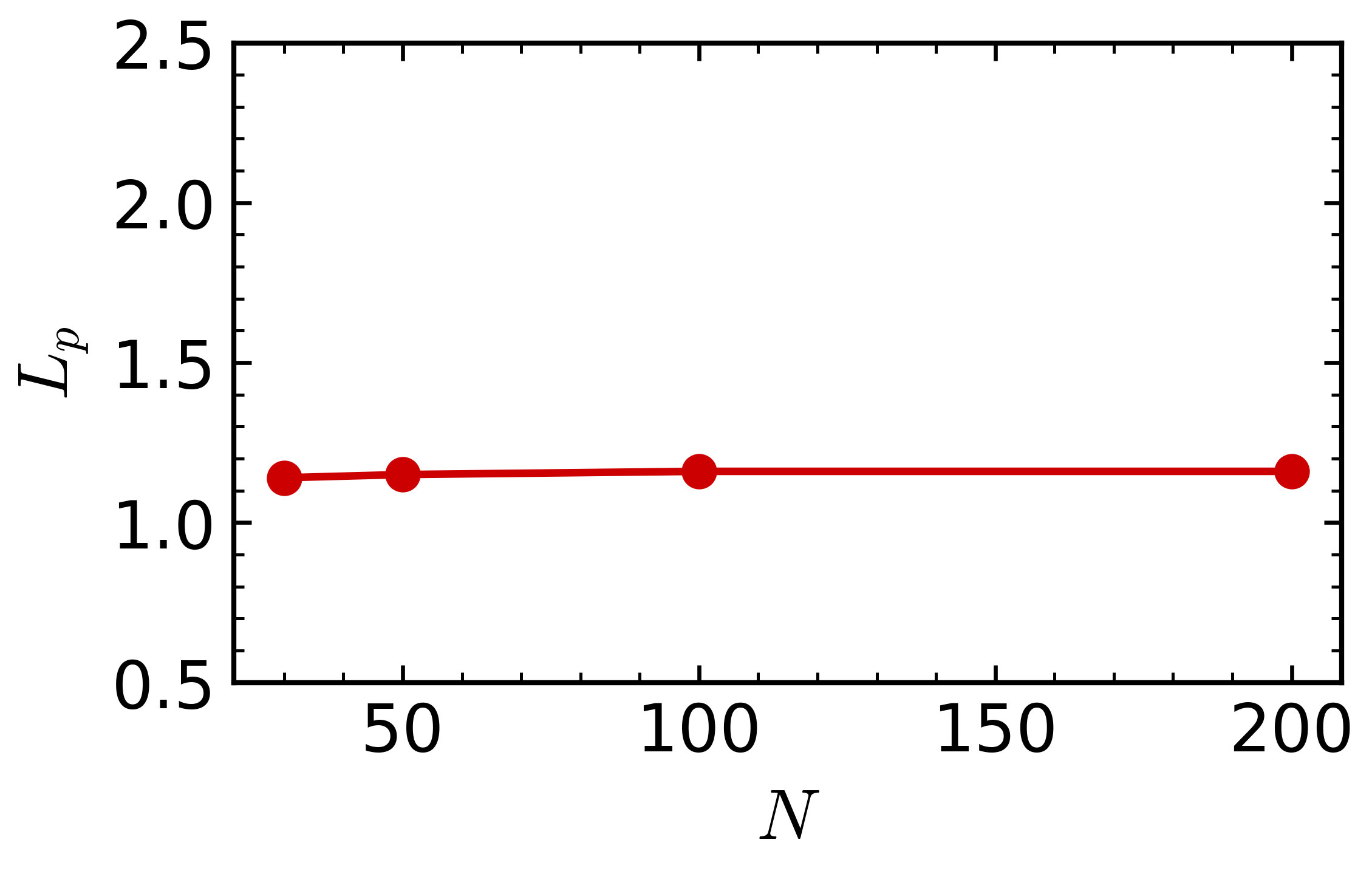}
    \caption{(Left) Flory's Characteristic Ratio $C_\infty$ vs chain length $N$. (Right) Persistence length $L_p$ vs chain length $N$.}
    \label{fig:SI_flory}
\end{figure}

\begin{figure}[!htbp]
    \centering
    \includegraphics[width = 0.5\textwidth]{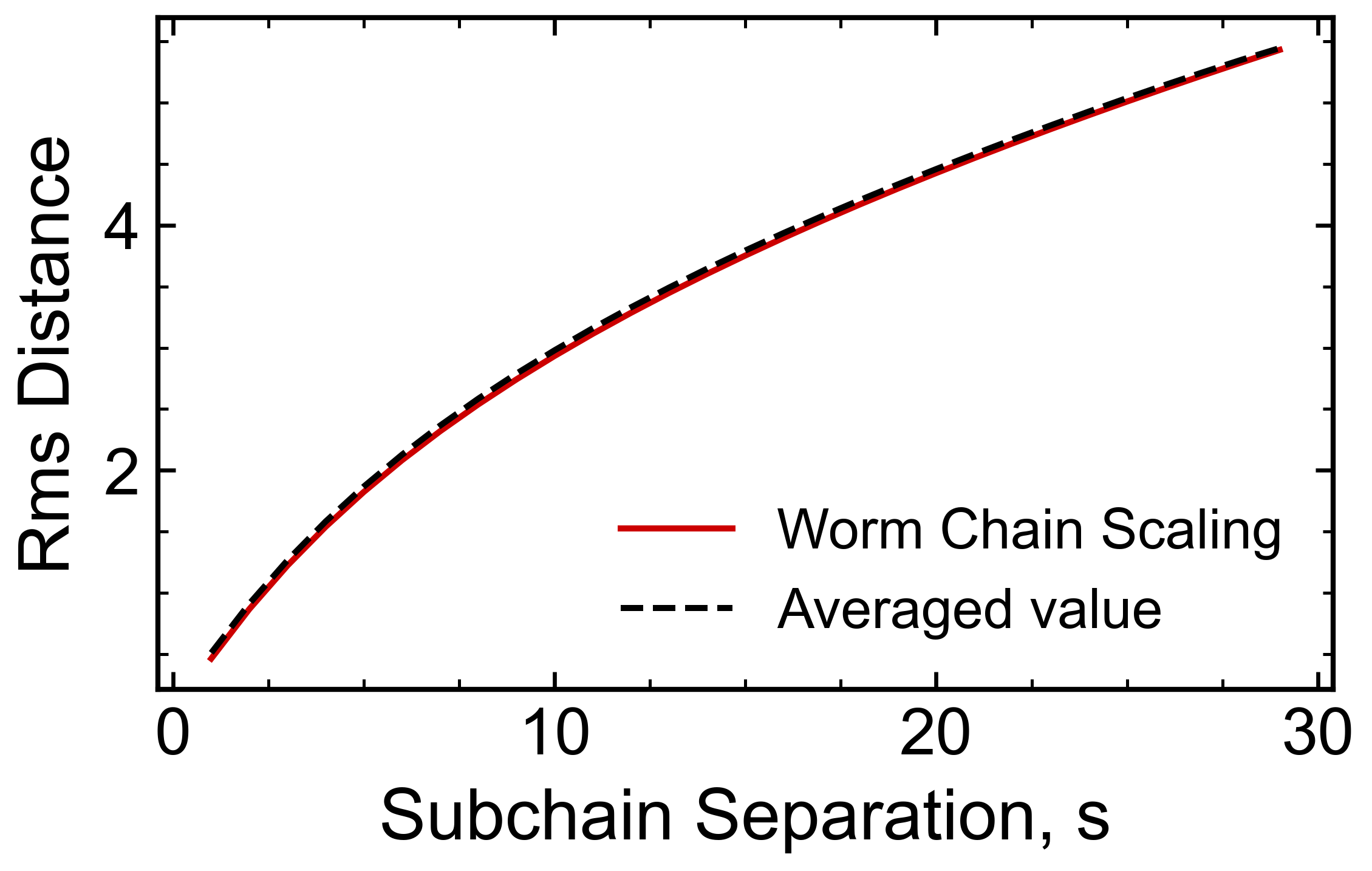} 
    \caption{Root mean squared distance vs. subchain separation $s$ of individual monomers connected by bonds in a melt. The measurement of one melt with $N = 30$ has been included. The Worm-like Chain model scaling has been used for comparison, to ensure that our configuration has fully relaxed before simulating our melts.}
    \label{fig:SI_RMSID}
\end{figure}

Because of the short bond length of 0.2, the segmental polymer model is worm-like in character, as shown by a characteristic ratio $C_\infty =4.2$ and a persistence length $L_p=1.1$ (Fig.~\ref{fig:SI_flory}). To ensure sufficiently equilibrated melts, we measured the Root Mean Squared Distance of chain segments \cite{Auhl_equilibration,sliozberg_equilibration} within a chain, and compared it to the expected worm-like scaling (Fig.\ref{fig:SI_RMSID}). All chain lengths followed the expected scaling. 

\begin{figure}[!htbp]
    \centering
\includegraphics{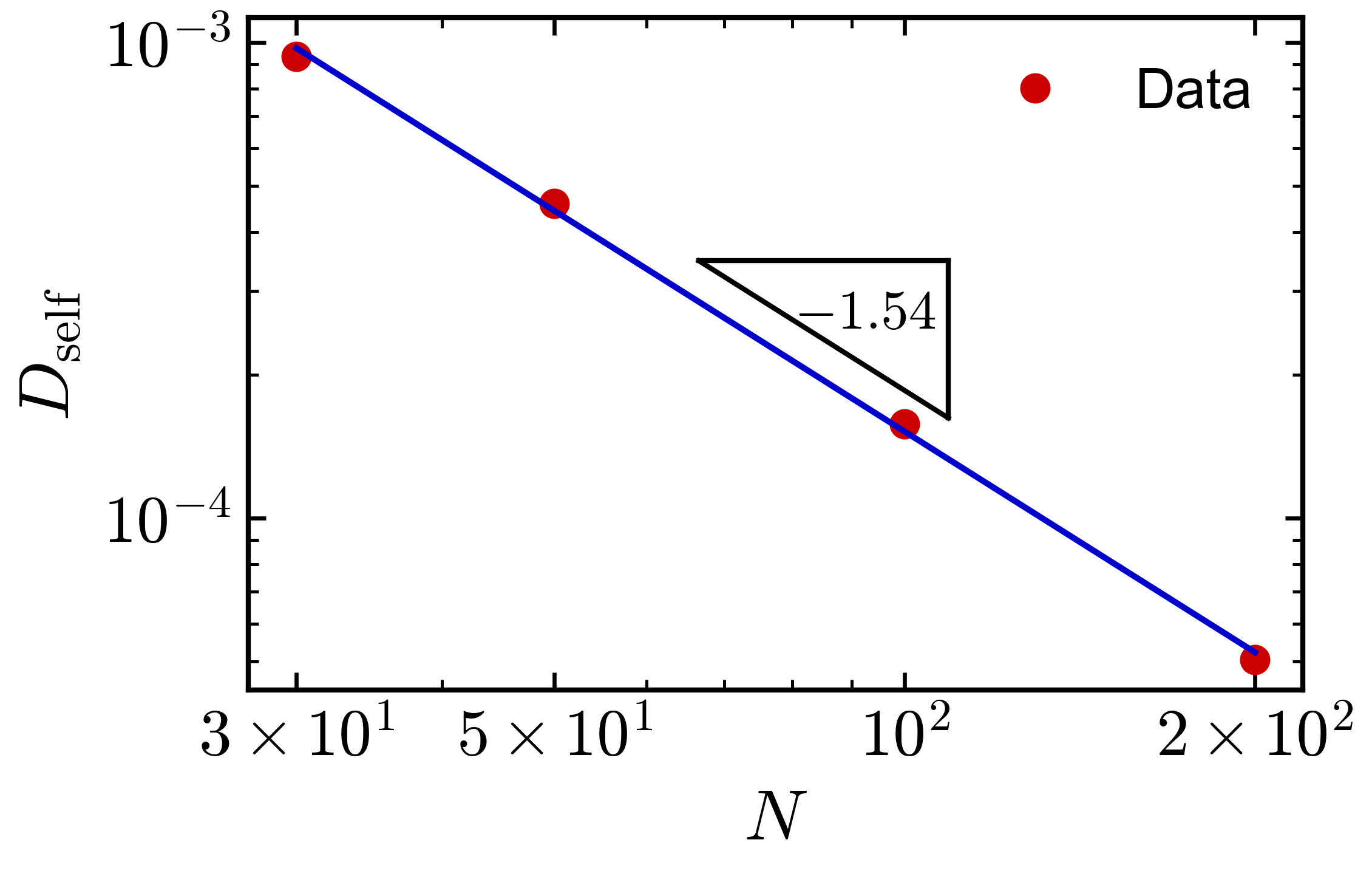}
\includegraphics{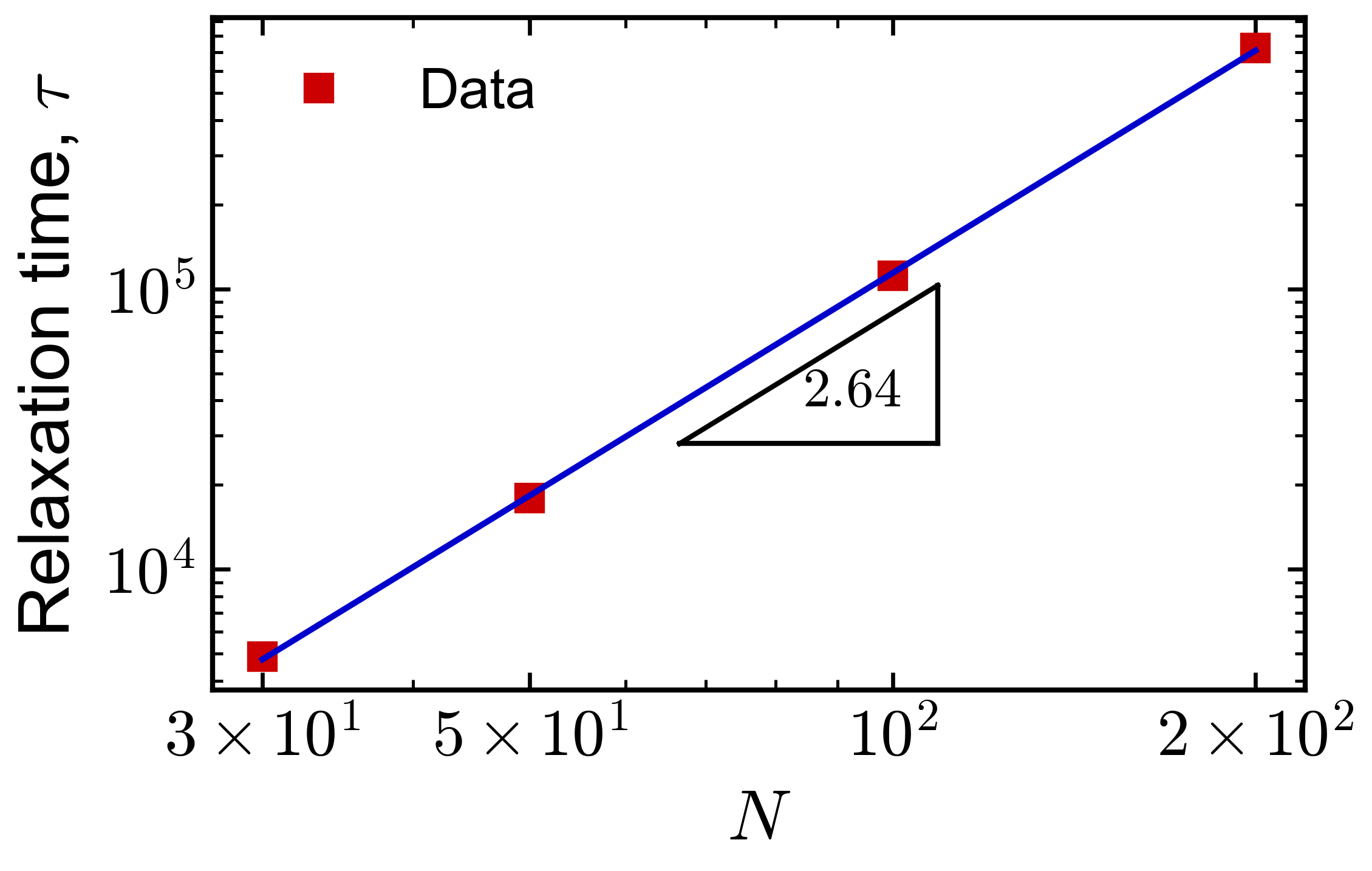}
    \caption{(Top) Self diffusion coefficient $D_\text{eff}$ scaling vs chain length $N$. (Bottom) Relaxation time $\tau$ scaling vs chain length$N$.}
    \label{fig:SI_diffusion_scaling}
\end{figure}

For determining the Weissenberg number Wi for the pre-shear step of the processing protocol, the diffusion coefficient $D_\text{eff}$ was measured from the mean-squared displacement and converted into a relaxation time $\tau$, shown in Fig.~\ref{fig:SI_diffusion_scaling}. 

The diffusion coefficient $D_\text{eff}$ scales with $N^{-1.5}$, which is between the unentangled Rouse $N^{-1}$, and entangled, reptation $N^{-2}$ scaling, consistent with the observation of a slight slowdown in the MSD for longer chains. Similarly, the relaxation time $\tau$ followed a scaling $\tau \sim N^{2.6}$, which is between Rouse $N^2$ and reptation $N^3$ scaling. We do not have enough chain lengths to clearly identify two distinct regimes.

\FloatBarrier
\section{Nematic Orientation and Flow-Induced Alignment}

To quantify molecular alignment during shear and quenching, we computed the global nematic order parameter $S$ using the \texttt{freud.order.Nematic} implementation.\cite{freud} Bond vectors between consecutive beads were treated as unit orientation vectors $\{\mathbf u_i\}$, from which the nematic tensor 
\[
\mathbf Q = \langle \mathbf u \otimes \mathbf u \rangle - \frac{1}{3}\mathbf I
\]
was constructed over all bonds in the system. The nematic order parameter was defined as $S = \frac{3}{2}\lambda_{\max}(\mathbf Q)$, where $\lambda_{\max}$ is the largest eigenvalue, and the corresponding eigenvector defines the global nematic director $\mathbf n$. The alignment is therefore measured relative to the emergent principal director of the system rather than with respect to a fixed laboratory axis. Under shear, this director coincides with the flow direction.

\begin{figure}[h]
    \includegraphics[width=0.8\linewidth]{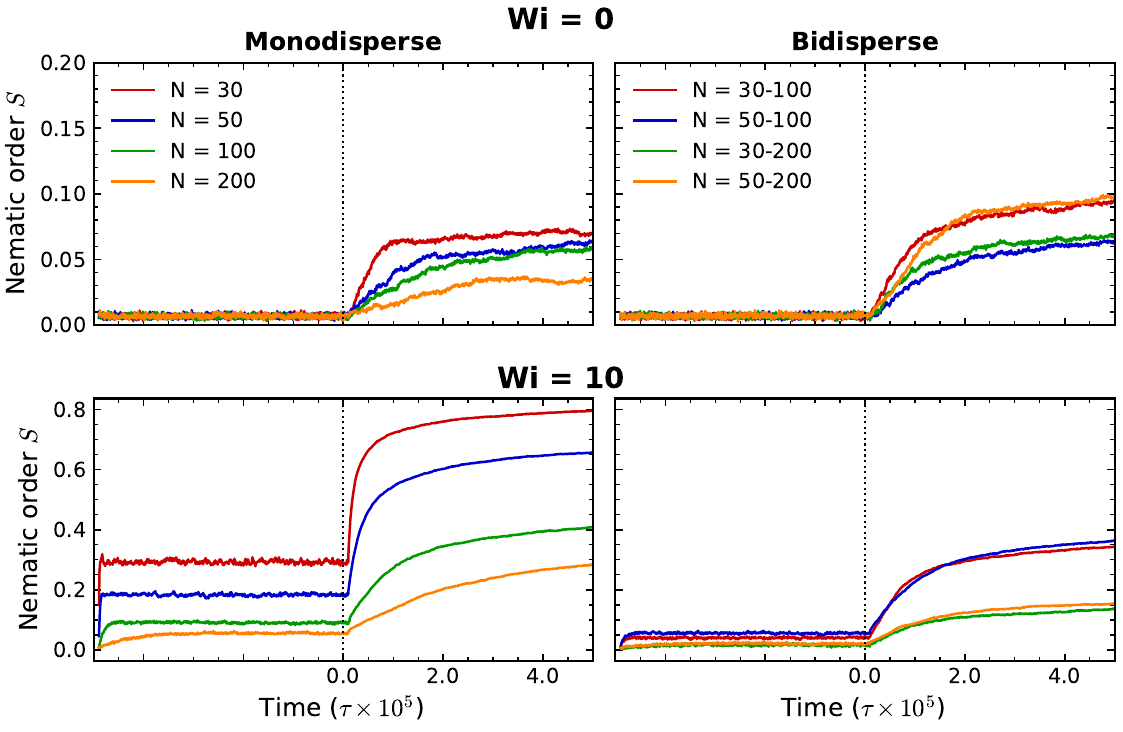}
    \caption{Time evolution of the global nematic order parameter $S$ for monodisperse (left) and bidisperse (right) melts under quiescent conditions (top, $\mathrm{Wi}=0$) and strong shear (bottom, $\mathrm{Wi}=10$). The vertical dotted line marks the termination of shear and the onset of the quench.}
    \label{fig:SI_nematic_4panel}
\end{figure}

Figure~\ref{fig:SI_nematic_4panel} shows the time evolution of $S$ for monodisperse and bidisperse melts under quiescent and sheared conditions. At $\mathrm{Wi}=0$, melts remain essentially isotropic prior to the quench, with $S \approx 0$ for all compositions. Under shear ($\mathrm{Wi}=10$), $S$ rises rapidly and reaches a plateau whose magnitude decreases with increasing chain length in the monodisperse series and is reduced in bidisperse mixtures containing the longest component, reflecting the slower orientational response of longer, more constrained chains.

Following the quench (right of the vertical line), $S$ increases sharply in all systems, signaling crystallization-induced alignment. The increase is most pronounced for short-chain monodisperse melts subjected to shear, where pre-existing flow alignment is amplified as crystalline domains grow along the established director. In systems containing longer chains, the post-quench enhancement of $S$ is weaker, consistent with reduced chain mobility and slower structural reorganization during crystallization.

\FloatBarrier
\subsection{Global Ordering: Structure Factor Analysis}

\begin{figure}[h]
    \includegraphics[width=0.6\linewidth]{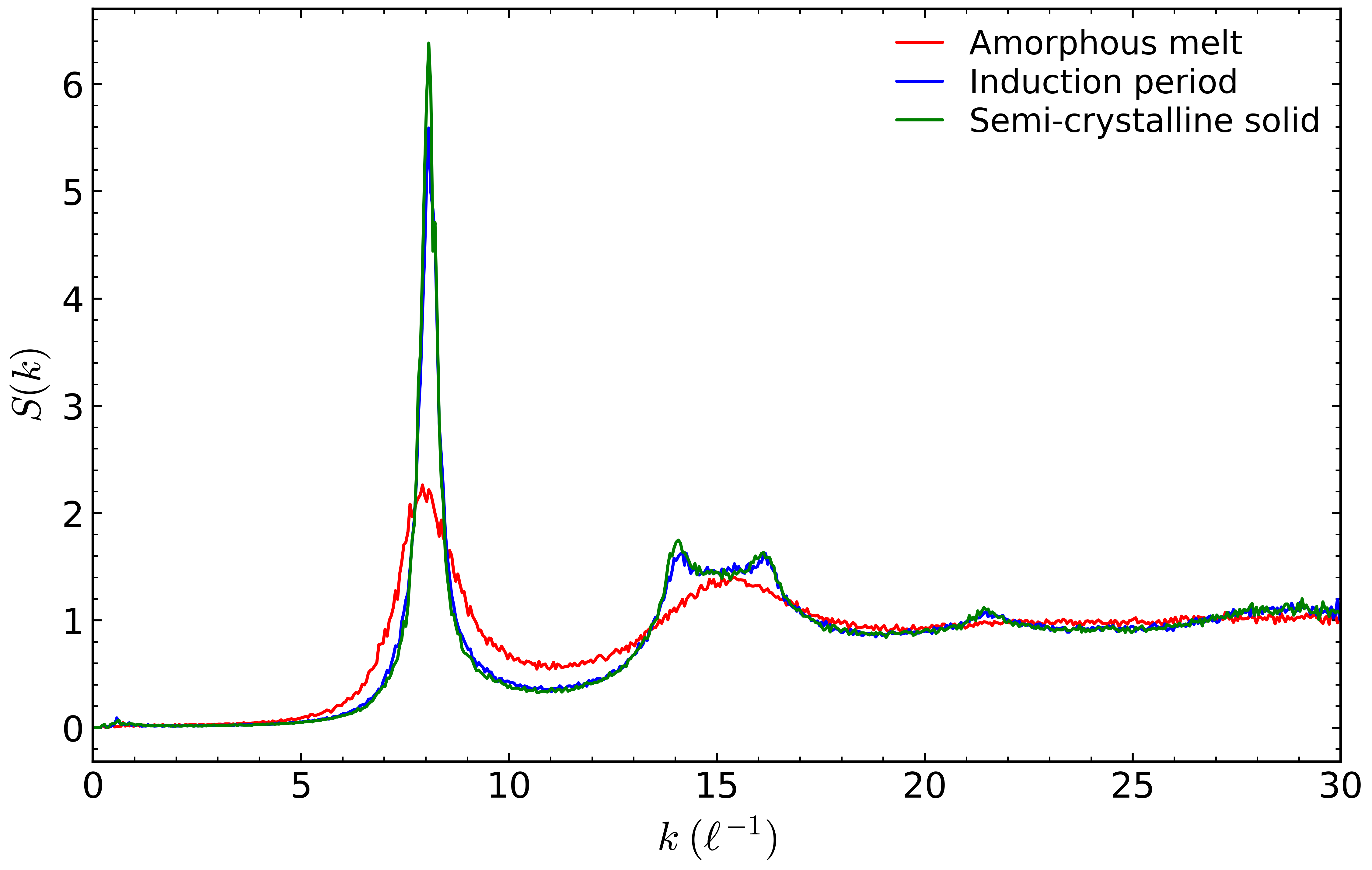}
    \caption{}
    \label{fig:SI_structure_factor}
\end{figure}

To characterize long-range periodicity and identify early-stage ordering in the melt, we computed the static structure factor following Li et al.\cite{li_unifying_2021},
\begin{equation}
S(\mathbf{q}) = \frac{1}{\sum_{i=1}^{N} a_i^2}
\left\langle \left| \sum_{i=1}^{N} a_i e^{i \mathbf{q}\cdot\mathbf{r}_i} \right|^2 \right\rangle.
\end{equation}
The emergence and growth of sharp peaks in $S(\mathbf{q})$ signal the transition from amorphous to crystalline order.
We computed $S(\mathbf{q})$ at fixed intervals throughout the quench to track the evolution of orientational order and lamellar spacing. The result for a N = 30 melt is plotted in Figure~\ref{fig:SI_structure_factor}.
Starting from the amorphous melt plotted with red, we gradually see the development of peaks in the Structure Factor that grow as the amorphous melt is subjected to pre-shear treatment (blue), which become stable when the amorphous melt has transitioned to a semi-crystalline solid (green). Sharper peaks of the Structure Factor correspond to higher crystallinity states\cite{Triandafilidi_2015}.

\begin{figure}[h]
    \includegraphics[width=\linewidth]{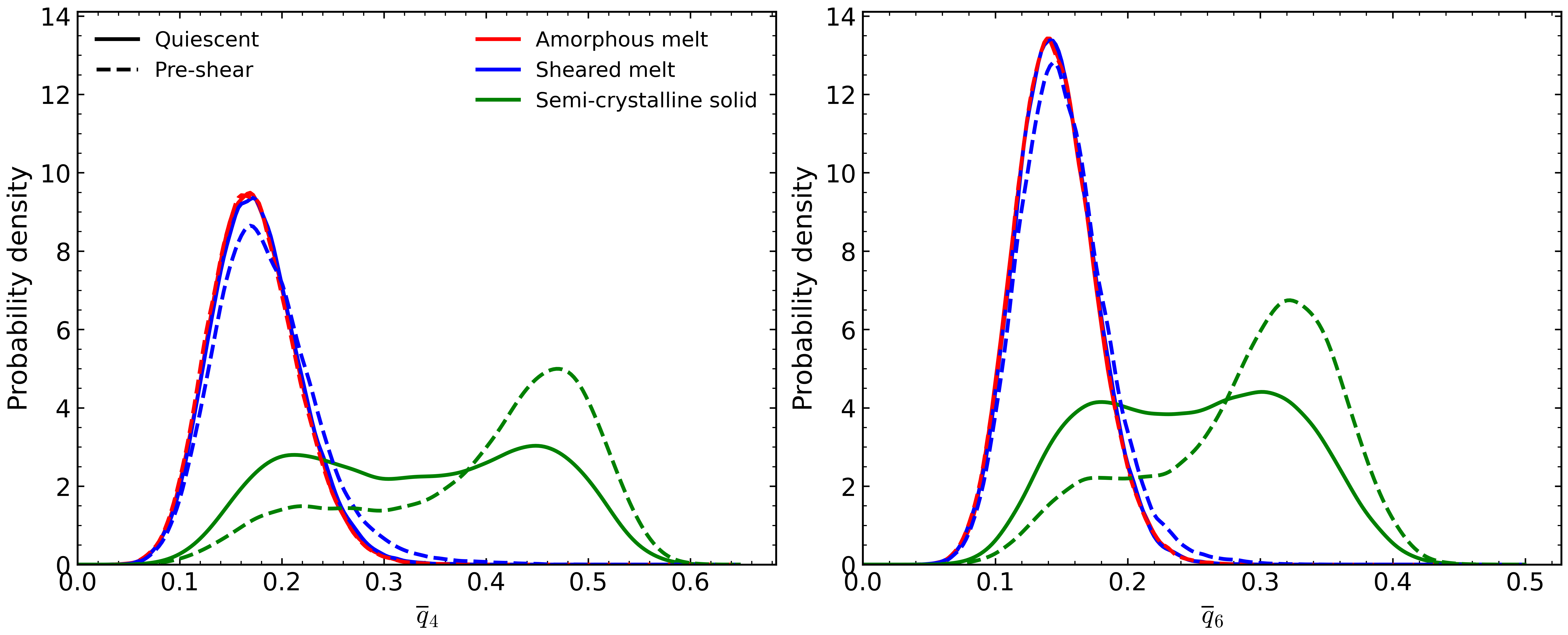}
    \caption{Steinhardt order parameters evolution $\bar q_4$ and $\bar q_6$ of the N = 30 melt at the amorphous, pre-shear, and crystalline state.}
    \label{fig:SI_histograms_q4_q6}
\end{figure}

To determine thresholds for the crystallinity analysis, we determined histograms of the $\bar q_6$ and $\bar q_4$ Steinhardt bond order parameter distributions. From Fig.~\ref{fig:SI_histograms_q4_q6} it is clear that the crystalline regions had values of both $\bar q_6$ and $\bar q_4$ above $\sim 0.25$ and amorphous melts had values below. In the cooled, semi-crystalline solids, the histograms showed two clear peaks, with a minimum at $ \bar q_6 \approx 0.25$, which we then used as threshold for identifying crystal domains and grains. Shearing shifted the amorphous distribution to slightly higher values, however, no second peak appeared, indicating that no true crystal nucleation was observed in the pre-sheared melts. 

\FloatBarrier
\subsection{Precursors}

\begin{figure}[h]
    \includegraphics[width=0.9\linewidth]{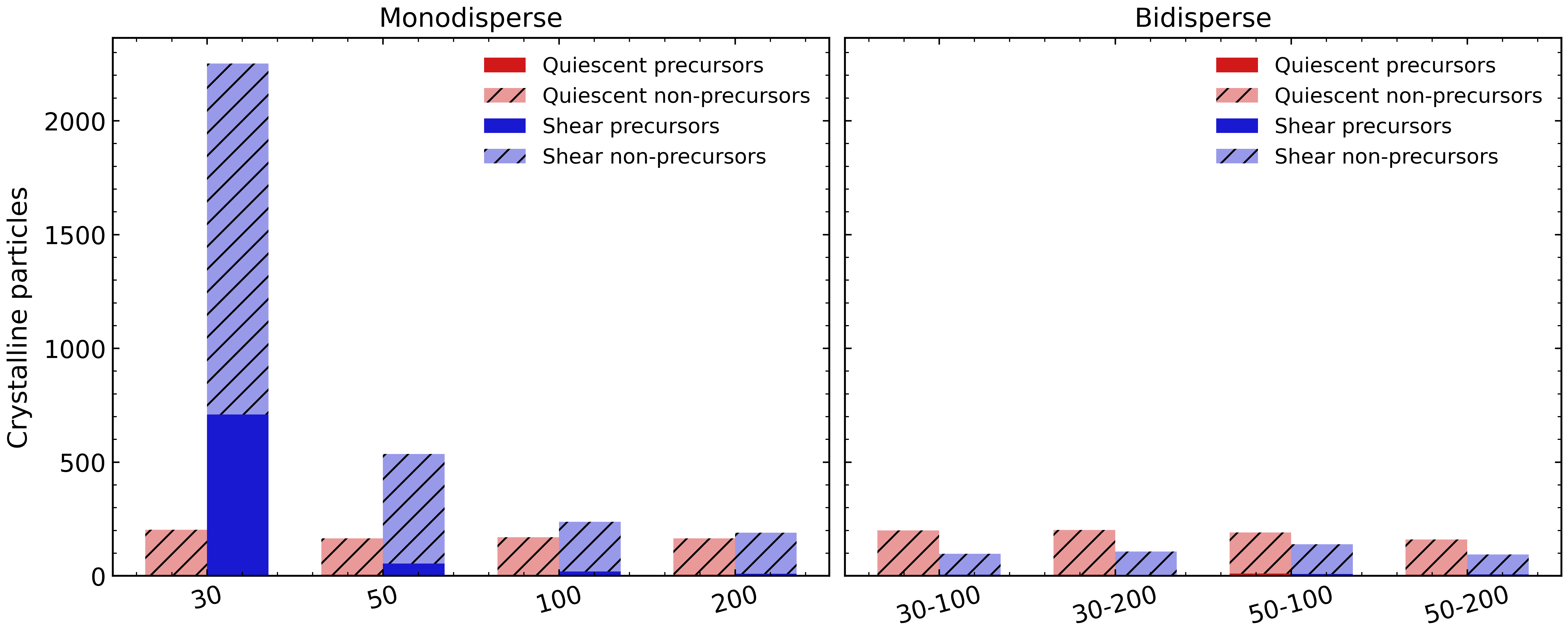}
    \caption{Number of crystalline particles in the first frame after the quench, identified by whether (solid part of the bar) or not (hatched part of the bar) they were identified as precursors at the end of pre-shear. The total number of monomers in each system was $81000$, e.g., $2000$ crystalline particles equate to $2.5\%$.}
    \label{fig:SI_precursor_conversion}
\end{figure}

To probe whether locally ordered or nematic liquid environments are associated with early
crystal formation, we classified particles at the end of the liquid trajectory
according to a bond-orientational criterion based on Steinhardt $q_4$ and $q_6$
(order parameters computed from Voronoi neighbors): particles with
$q_4 > 0.2$ and $q_6 > 0.2$ were retained as \emph{candidate} precursors, without
implying that they are true nucleation sites.
We then followed the quench trajectory and recorded, among those candidates,
how many adopt a crystalline label within the first frames after the quench. Overall, only the short monodisperse system showed a appreciable fraction of crystalline particles this early on after the quench, around $2.7\%$. 
The resulting conversion fractions for each system, summarized in Fig.~\ref{fig:SI_precursor_conversion},
suggest a systematic dependence on chain length in the pre-sheared monodisperse
melts: for the shortest chains ($N=30$) a sizable share of pre-cursor candidates
crystallizes early, and that share decreases rapidly as $N$ increases.
By contrast, the bidisperse compositions do not show the same monotonic trend,
and the corresponding fractions remain comparatively small, indicating that the
link between this local-order proxy and immediate post-quench crystallization
is much weaker to non-exisitent in those mixtures.

\FloatBarrier
\section{Loop and Tie chain time evolutions}
\begin{figure}
    \includegraphics[width=0.8\linewidth]{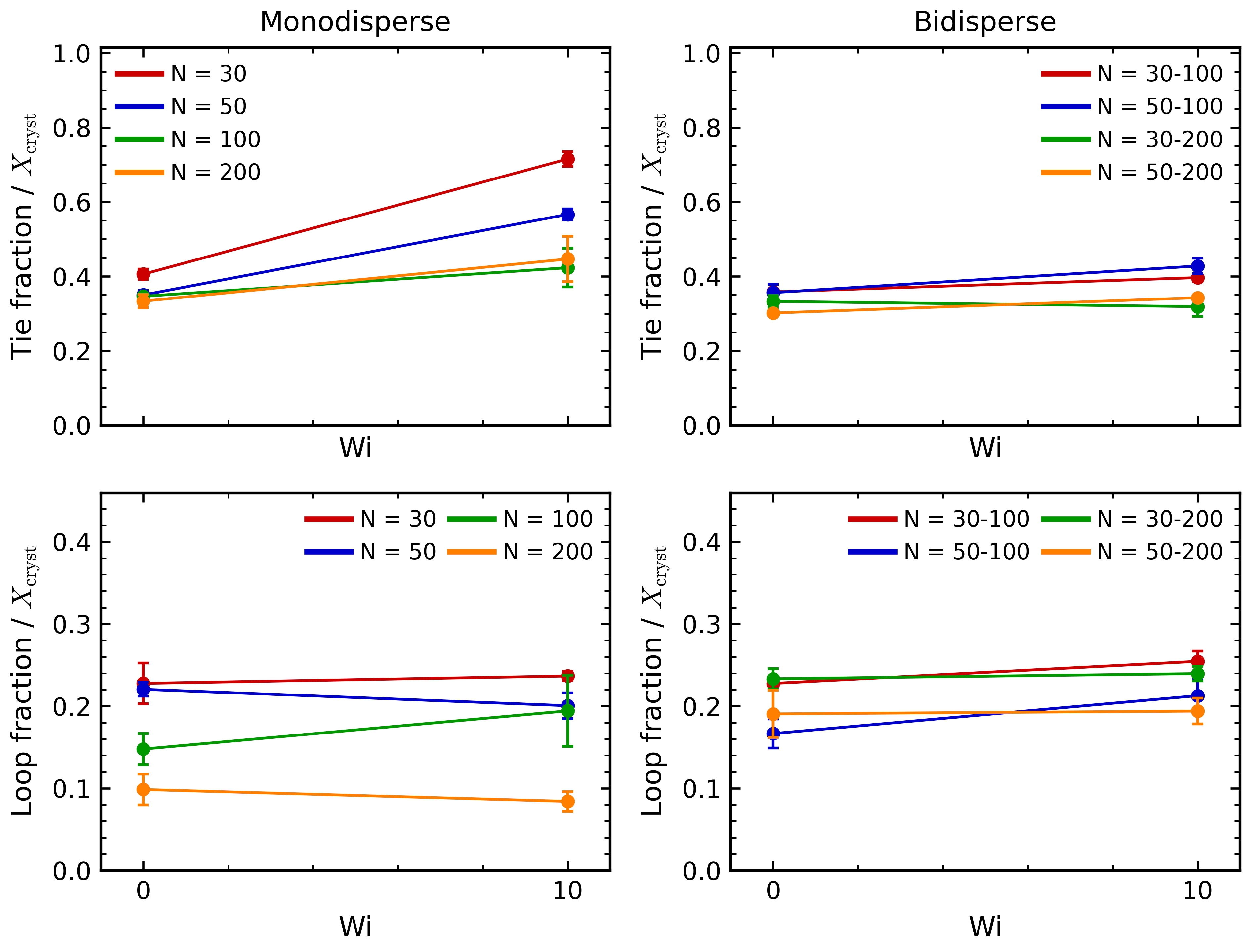}
    \caption{Fraction of tie and loop chains over total number of chains in Monodisperse and bidisperse melts with varying Weissenberg number, normalized by the average crystallinity of each melt.}
    \label{fig:SI_normalized_tie_loop_end_values}
\end{figure}

 In Fig.~\ref{fig:SI_normalized_tie_loop_end_values} the same data as in the main manuscript is shown, the fraction of loop and tie chains at the end of the simulation. Here, those values were normalized by the fraction of the system that is crystalline to confirm that the observed trends are not an effect of the fact that some melts were more or less crystalline. The trends discussed in the main manuscript persist.
 
\begin{figure}
    \includegraphics[width=0.8\linewidth]{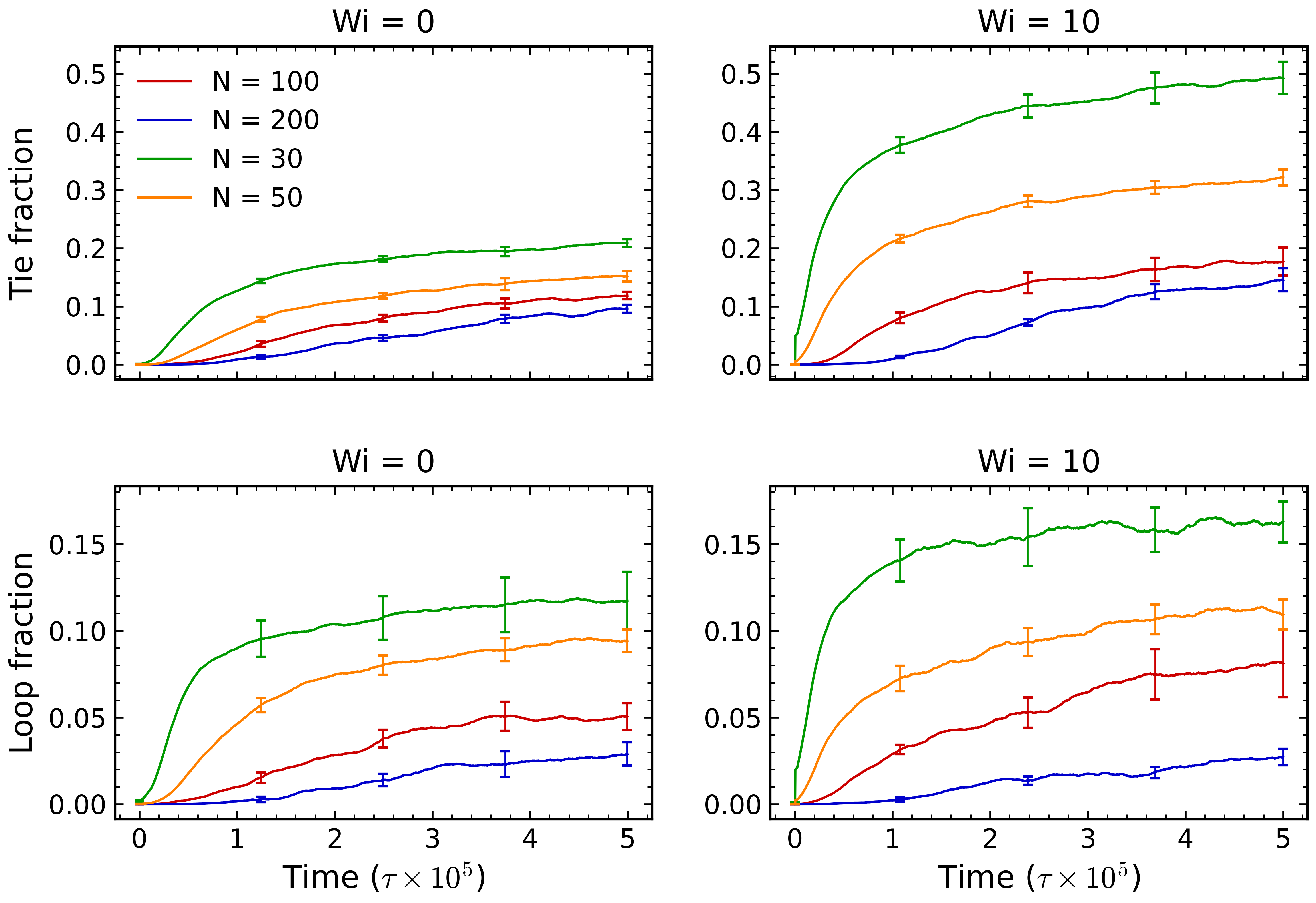}
    \caption{Tie and loop chain structures and their time-series evolution in monodisperse melts at quiescent conditions (Wi = 0) or under an imposed shear flow field (Wi = 10).}
    \label{fig:SI_monodisperse_topology}
\end{figure}

\begin{figure}
    \includegraphics[width=0.8\linewidth]{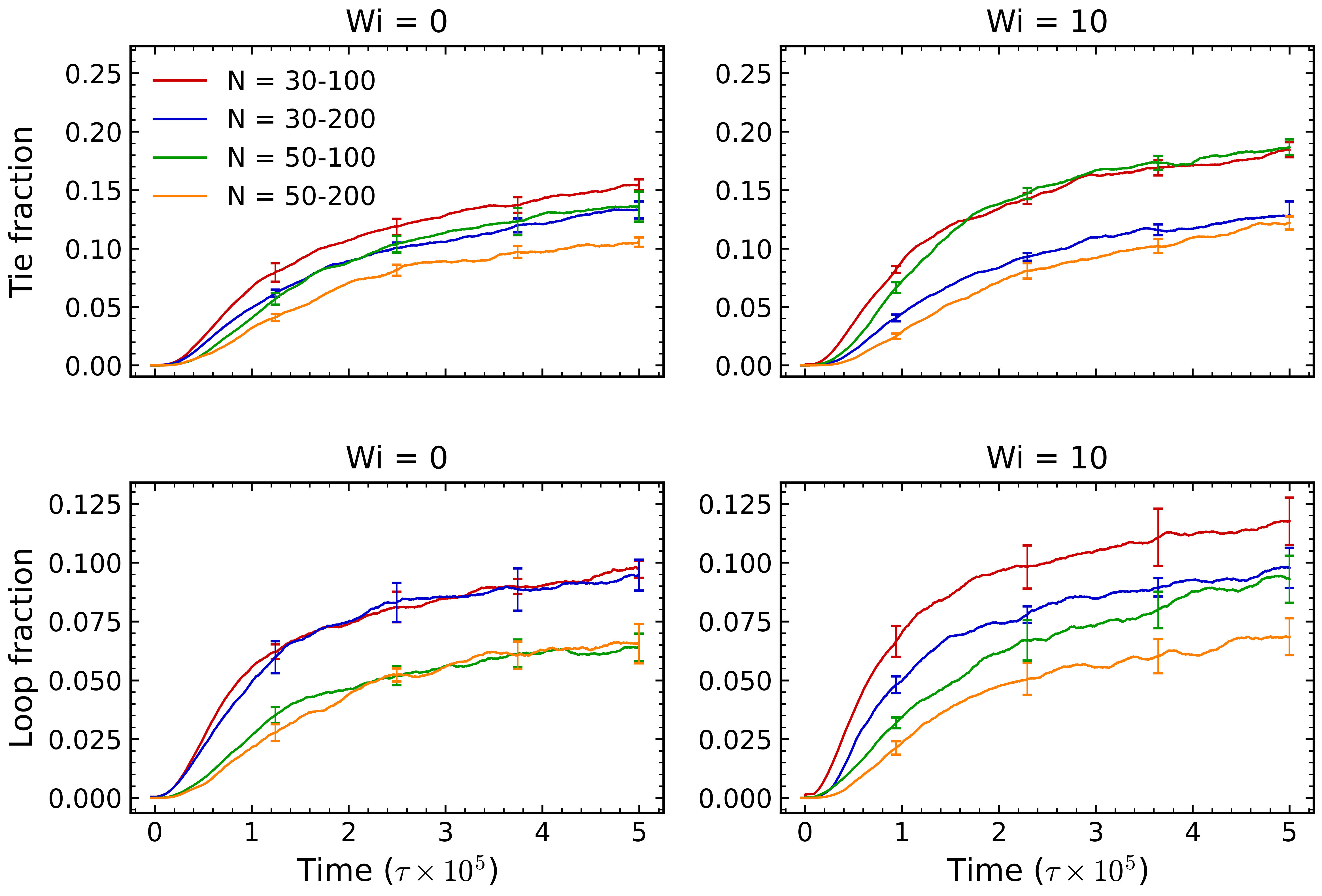}
    \caption{Tie and loop chain structures and their time-series evolution in bidisperse melts at quiescent conditions (Wi = 0) or under an imposed shear flow field (Wi = 10).}
    \label{fig:SI_bidisperse_topology}
\end{figure}

As expected and shown in Fig.~\ref{fig:SI_monodisperse_topology} and Fig.~\ref{fig:SI_bidisperse_topology}, both tie and loop chain fractions increased over time. Both loop and tie fractions were also enhanced by the presence of pre-shear treatment in the bidisperse blends, but to a lesser extent. 

\begin{figure}
    \includegraphics[width=1.0\linewidth]{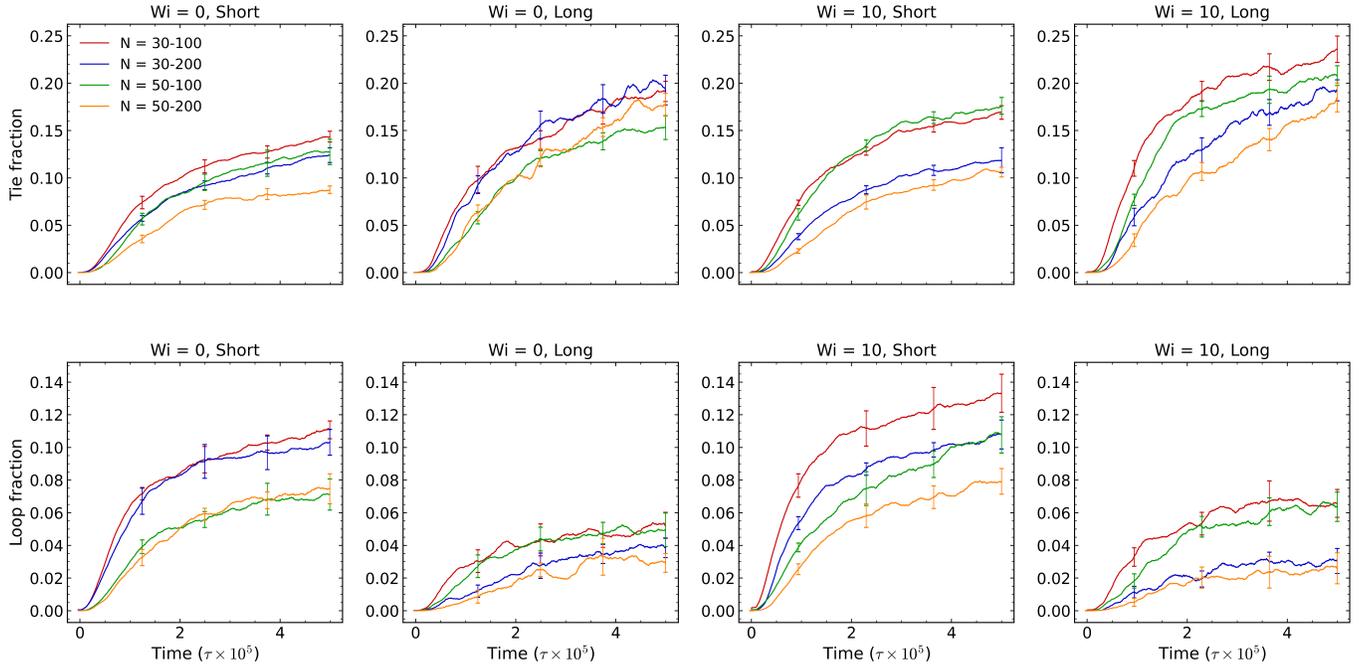}

    \caption{Tie and loop chain structures distribution in short and long chains of bidisperse melts and their time-series evolution at quiescent conditions (Wi = 0) or under an imposed shear flow field (Wi = 10).}
    \label{fig:SI_tie_loop_short_long}
\end{figure}

In Fig.~\ref{fig:SI_tie_loop_short_long}, we analyzed the fraction of loop and tie chains in the bidisperse blends, divided by whether they were the short or long component in the blend. The top row shows the tie chains, the bottom row the loop chains. On the left, the blends without shear are shown, whereas the pre-shear blends are on the right. Under quiescent condition, the long chains formed slightly more tie chains than the short, and slightly less loop chains than the short chains. With pre-shear, this trend persisted, with a minor increase of the long chain loop fraction.

\FloatBarrier
\subsection{Gyration Tensor calculation}

The gyration tensor was calculated using the  definition: 

\begin{equation}
G_{\alpha\beta}
=
\frac{1}{N_c}
\sum_{i=1}^{N_c}
\left(r_{i,\alpha}-R_{\mathrm{cm},\alpha}\right)
\left(r_{i,\beta}-R_{\mathrm{cm},\beta}\right),
\label{eq:gyration_tensor}
\end{equation}
where $\alpha,\beta \in \{x,y,z\}$ denote Cartesian components.

The gyration tensor was diagonalized to obtain its principal axes and principal moments through
\begin{equation}
\mathbf{G}\,\mathbf{v}_i = \lambda_i \mathbf{v}_i,
\qquad i=1,2,3,
\label{eq:eigenproblem}
\end{equation}
where \(\lambda_i\) are the eigenvalues of \(\mathbf{G}\) and \(\mathbf{v}_i\) are the corresponding orthonormal eigenvectors. The eigenvalues were sorted in ascending order,
\begin{equation}
\lambda_1 \le \lambda_2 \le \lambda_3.
\label{eq:eigenvalue_order}
\end{equation}

The squared radius of gyration is given by the trace of the gyration tensor,
\begin{equation}
R_g^2 = \mathrm{Tr}\,\mathbf{G} = \lambda_1 + \lambda_2 + \lambda_3.
\label{eq:rg2}
\end{equation}

The cluster asphericity was defined as
\begin{equation}
AS =
\lambda_3 - \frac{1}{2}\left(\lambda_1+\lambda_2\right),
\label{eq:asphericity}
\end{equation}
and its normalized form was computed as
\begin{equation}
\frac{bAS}{R_g^2}
=
\frac{\lambda_3 - \frac{1}{2}\left(\lambda_1+\lambda_2\right)}
{\lambda_1+\lambda_2+\lambda_3}.
\label{eq:asphericity_norm}
\end{equation}

The cluster prolateness was defined as
\begin{equation}
p =
\left\langle
\frac{
(2\lambda_1-\lambda_2-\lambda_3)
(2\lambda_2-\lambda_1-\lambda_3)
(2\lambda_3-\lambda_1-\lambda_2)
}{
2\left(
\lambda_1^2+\lambda_2^2+\lambda_3^2
-\lambda_1\lambda_2-\lambda_2\lambda_3-\lambda_1\lambda_3
\right)^{3/2}
}
\right\rangle
\label{eq:prolateness}
\end{equation}

The eigenvectors of the gyration tensor define the principal directions of the cluster, while the corresponding eigenvalues quantify the spatial extent along those directions.

\FloatBarrier
\section{Model - Methods}

\subsection{SLLOD Equations of Motion for Homogeneous Shear Flow}

We implement homogeneous planar shear flow in a fork of HOOMD-blue by extending the constant-volume two-step integrator to include the SLLOD equations of motion under Lees--Edwards (triclinic) boundary conditions.\cite{todd2017nonequilibrium,daivis_simple_2006,edwards_validation_2006} The code is available at \url{https://github.com/stattlab/hoomd-blue/tree/feature/sllod}.
This formulation enables nonequilibrium molecular dynamics (NEMD) simulations of homogeneous shear while preserving periodicity and Galilean invariance.

\subsection{Equations of Motion}

Let the streaming field be defined as
\[
\vec{u}(\vec{r}) = \dot{\gamma}\, y\, \hat{\vec{x}},
\]
with the velocity gradient tensor $\boldsymbol{\kappa} = \nabla \vec{u}$, for which $\kappa_{xy} = \dot{\gamma}$ under simple shear.  
We define the peculiar velocity and momentum as
\[
\vec{c}_i = \vec{v}_i - \vec{u}(\vec{r}_i), \qquad \vec{p}_i = m_i\,\vec{c}_i,
\]
where $m_i$, $\vec{r}_i$, and $\vec{v}_i$ denote the mass, position, and velocity of particle $i$, respectively.  
The SLLOD equations of motion are then given by
\begin{equation}
  \dot{\vec{r}}_i = \frac{\vec{p}_i}{m_i} + \boldsymbol{\kappa}\,\vec{r}_i,
  \qquad
  \dot{\vec{p}}_i = \vec{F}_i^{\phi} - \boldsymbol{\kappa}\,\vec{p}_i,
  \label{eq:sllod}
\end{equation}
where $\vec{F}_i^{\phi}$ is the total interparticle force on particle $i$.  
For simple shear, the streaming velocity is $\vec{u} = (\dot{\gamma}y, 0, 0)$.  
These equations are formally equivalent to Newton’s equations of motion for a fluid subject to an external shear field.\cite{soddemann2003dissipative}

\FloatBarrier
\subsection{Boundary Conditions and Box Evolution}

Lees--Edwards boundary conditions are employed to maintain continuous particle trajectories during shear deformation.  
The triclinic simulation box is sheared by updating the global tilt factor $xy$ each time step:
\[
xy(t+\Delta t) = xy(t) + \dot{\gamma}\,\Delta t.
\]
When $|xy| > 0.5$, the box is remapped into the interval $[-0.5,0.5)$ and all particle images are updated consistently to preserve trajectory continuity.\cite{todd2017nonequilibrium}

More generally, the evolution of the box lattice vectors $\mathbf{L}_k = (L_{kx}, L_{ky}, L_{kz})$ follows
\begin{equation}
\dot{\mathbf{L}}_k(t) = \mathbf{L}_k(t) \cdot \boldsymbol{\kappa},
\label{eq:box_evolution}
\end{equation}
which, for planar shear flow, reduces to
\begin{align*}
\dot{L}_{kx}(t) &= L_{ky}(t)\,\dot{\gamma}, \quad
\dot{L}_{ky}(t) = 0, \quad
\dot{L}_{kz}(t) = 0.
\end{align*}

If particles cross the top or bottom boundaries in the $y$-direction, they are wrapped back into the box through the opposite face and displaced in the flow direction by $\pm L_y\,\dot{\gamma}\,\Delta t$.  
The tilt angle $\theta$ increases until it reaches $45^\circ$, at which point the box is remapped to its original cubic orientation, and particle images are shifted accordingly.


\FloatBarrier
\subsection{Thermostatting}

In nonequilibrium steady-state simulations, temperature control must be applied to \textit{peculiar} velocities only, ensuring profile-unbiased thermostating and preservation of Galilean invariance.\cite{yong2013thermostats}  
In our implementation, the streaming component $\vec{u}(\vec{r})$ is subtracted before applying the thermostat and re-added afterward.  
We employ a Nosé--Hoover thermostat acting on the peculiar velocities:
\begin{equation}
    \vec{v}_i' = \vec{v}_i - \vec{u}(\vec{r}_i),
\end{equation}
which maintains the target temperature while leaving the macroscopic shear profile unaffected.

For DPD-style thermostats, the random and dissipative forces satisfy the fluctuation–dissipation relation:
\begin{equation}
\langle W_i(t)\,W_j(t') \rangle = \delta_{ij}\,\delta(t-t')\,6\,\Gamma k_B T,
\end{equation}
where the friction coefficient $\Gamma$ was set to 1 in all production runs, yielding a steady and spatially uniform temperature profile without artificial fluctuations.\cite{yong2013thermostats}

A separate class was implemented to compute thermodynamic quantities using peculiar velocities.  
Kinetic temperature, pressure tensor, and energies are computed based on the peculiar, rather than total, velocities, ensuring that the measured temperature corresponds to the thermal motion only.

\FloatBarrier
\subsection{Integrator Algorithm}

\begin{tcolorbox}[title={Algorithm S1: Two-step constant-$V$ SLLOD integrator (per time step $\Delta t$)},colback=white,colframe=black]
\begin{enumerate}\itemsep3pt
  \item \textbf{Box update:} $xy \leftarrow xy + \dot{\gamma}\,\Delta t$; if $|xy|>0.5$, remap $xy$ into $[-0.5,0.5)$ and adjust particle images (Lees--Edwards).
  \item \textbf{Half step (per particle):}
    \begin{enumerate}\itemsep2pt
      \item Remove flow: $\vec{v}\leftarrow \vec{v}-\vec{u}(\vec{r})$.
      \item Thermostat half-step on peculiar $\vec{v}$.
      \item Half-kick: $\vec{v}\leftarrow \vec{v}+\tfrac{1}{2}\vec{a}\,\Delta t$, with $\vec{a}=\vec{F}/m$.
      \item Drift: $\vec{r}\leftarrow \vec{r}+\vec{v}\,\Delta t$; apply Lees--Edwards wrapping (shear-consistent shifts).
      \item Add flow back: $\vec{v}\leftarrow \vec{v}+\vec{u}(\vec{r})$.
    \end{enumerate}
  \item \textbf{Forces:} rebuild neighbor lists and compute $\vec{F}$.
  \item \textbf{Second half step (per particle):}
    \begin{enumerate}\itemsep2pt
      \item Half-kick: $\vec{v}\leftarrow \vec{v}+\tfrac{1}{2}\vec{a}\,\Delta t$ (using updated forces).
      \item Remove flow; thermostat half-step on peculiar $\vec{v}$; add flow back.
    \end{enumerate}
\end{enumerate}
\end{tcolorbox}

\FloatBarrier
\subsection{Implementation (Summary)}

The CPU and CUDA code paths share identical update ordering.  
We list below the files containing substantive implementation logic (omitting trivial headers):

\begin{itemize}
  \item \texttt{TwoStepConstantVolumeSLLOD.cc} — Host-side integrator handling triclinic tilt updates, flow subtraction/addition around thermostatting, and Lees--Edwards wrapping.
  \item \texttt{TwoStepConstantVolumeSLLODGPU.cc},  \texttt{TwoStepConstantVolumeSLLODGPU.cu}, \\
  \texttt{TwoStepConstantVolumeSLLODGPU.cuh} — GPU implementations with CUDA kernels mirroring the host sequence.
\end{itemize}

\FloatBarrier
\subsection{Validation}

We assessed the implementation by comparing independent routes to calculating the viscosity at the same thermodynamic state points of a simple Lennard-Jones fluid.
In particular, we compared steady-state SLLOD estimates to equilibrium Green--Kubo viscosities~\cite{hess_berk} and to M\"uller--Plathe\cite{MP_reverse_original} (reverse perturbation) results, using LAMMPS trajectories extracted from our production logs.
We further compared LAMMPS SLLOD against HOOMD-blue SLLOD runs on CPU and GPU to confirm reproducibility across engines.
Across the temperatures surveyed, the methods are in quantitative agreement, with no systematic differences that could be attributed to the method of choice.

\begin{figure}
    \includegraphics[width=0.6\linewidth]{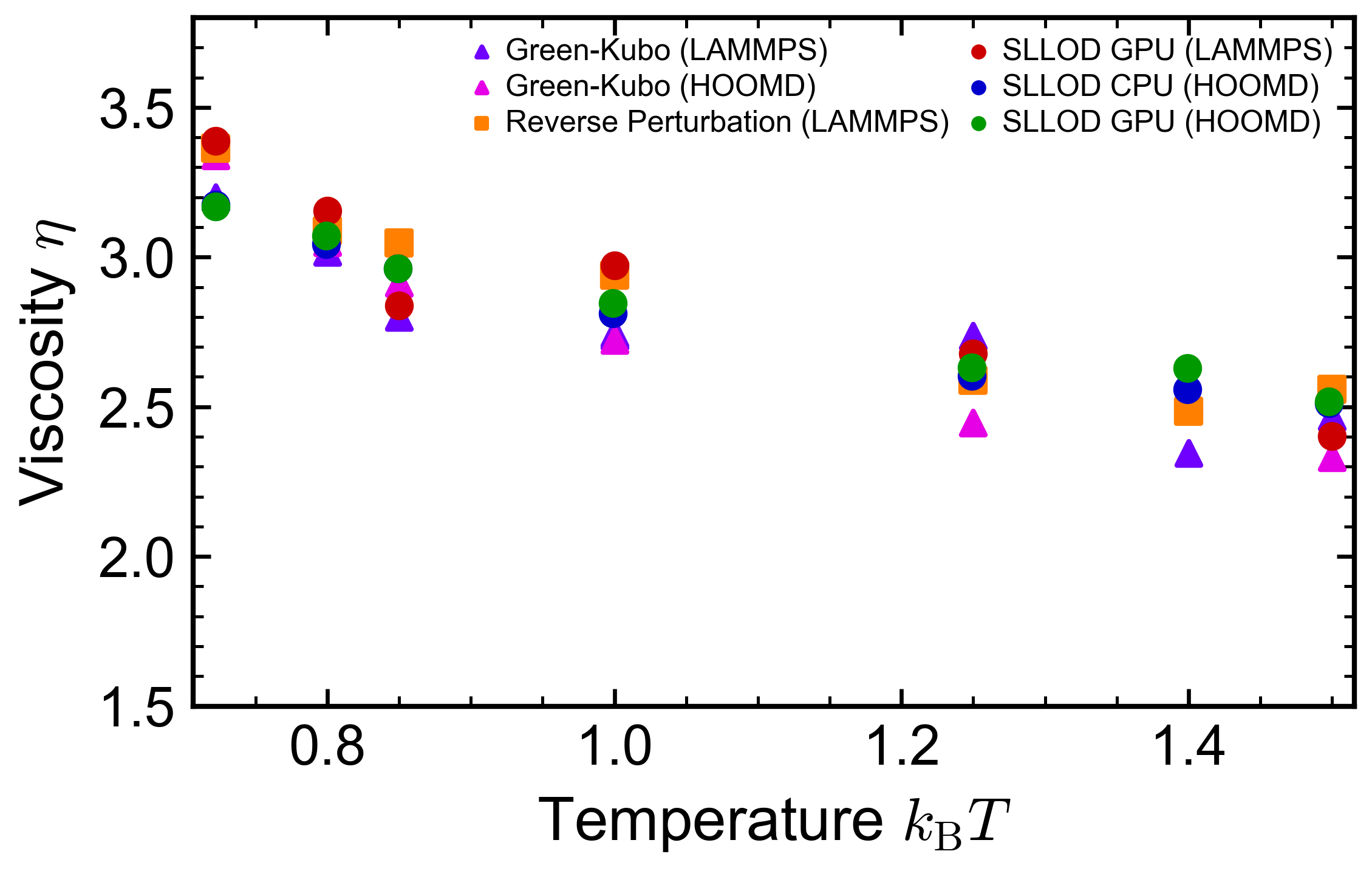}
    \caption{Viscosity $\eta$ of a LJ fluid versus $k_{\mathrm{B}}T$ at fixed bulk density $\rho = 0.8442/\ell^{3}$, comparing Green--Kubo (LAMMPS and HOOMD-blue), reverse-perturbation Müller--Plathe (LAMMPS), and steady SLLOD estimates (LAMMPS; HOOMD-blue SLLOD on CPU and GPU).}
    \label{fig:viscosity_validation}
\end{figure}


\FloatBarrier

\end{document}